\def\figsize{9.3cm}

\def\rn{}
\def\nn#1 #2{#2. #1}				
\def\nnn#1 #2 #3{#2. #3. #1}			
\def\nnnn#1 #2 #3 #4{#2. #3. #4 #1}		
\def\nnnnn#1 #2 #3 #4 #5{#2. #3. #4 #5. #1}	
\def\dualand{ and\hbox{ }}				
\def\multiand{, and\hbox{ }}				
\def\rf#1;#2;#3;#4;#5 {{\frenchspacing\par\rn#1, #3 {\bf #4}, #5 (#2). \par}}
\def\rg#1;#2;#3;#4;#5;#6 {{\frenchspacing\par\rn#1, #3 {\bf #4}, #5 (#2). \par}}
\def\rfproc#1;#2;#3;#4;#5;#6 {{\frenchspacing\par\rn#1 #2, in {\it #3}, ed. #4 (#5: #6)\par}}
\def\rfprocp#1;#2;#3;#4;#5;#6;#7 {{\frenchspacing\par\rn#1 #2, in {\it #3}, ed. #4 (#5: #6), p#7\par}}

\def\rg#1;#2;#3;#4;#5;#6 {\par\rn#1 #2, {\it #3}, {\bf #4}, #5 (``#6'') \par}
\def\rf#1;#2;#3;#4;#5 {\par\rn#1, {\it #3}, {\bf #4}, #5 (#2)\par}
\def\rfbook#1;#2;#3;#4;#5 {{\frenchspacing\par\rn#1, {\it #3} (#4: #5, #2)\par}}
\def\rfproc#1;#2;#3;#4;#5;#6 {{\frenchspacing\par\rn#1 #2, in {\it #3}, ed. #4 (#5: #6)\par}}
\def\rfprocp#1;#2;#3;#4;#5;#6;#7 {{\frenchspacing\par\rn#1 #2, in {\it #3}, ed. #4 (#5: #6), p#7\par}}
\def\rfprep#1;#2;#3 {{\par\frenchspacing\rn#1, #3 (#2)\par}}
\def\rfprepp#1;#2;#3 {{\par\rn#1 #2, #3\par}}

\def\Mpc{{\rm Mpc}}
\def\Gpc{{\rm Gpc}}

\def\MHz{{\rm MHz}}

\def\expec#1{\langle#1\rangle}

\def\etal{{\frenchspacing\it et al.}}
\def\ie{{\frenchspacing\it i.e.}}
\def\eg{{\frenchspacing\it e.g.}}
\def\etc{{\frenchspacing\it etc.}}

\def\beq#1{\begin{equation}\label{#1}}
\def\eeq{\end{equation}}
\def\beqa#1{\begin{eqnarray}\label{#1}}
\def\eeqa{\end{eqnarray}}
\def\eq#1{equation~(\ref{#1})}
\def\Eq#1{Equation~(\ref{#1})}
\def\eqn#1{~(\ref{#1})}

\def\fig#1{Figure~\ref{#1}}
\def\Fig#1{Figure~\ref{#1}}

\def\ScopeComparisonTable{1}
\def\ResolutionComparisonTable{2}

\def\Sec#1{Section~\ref{#1}}
\def\Sec#1{Section~\ref{#1}}

\def\spose#1{\hbox to 0pt{#1\hss}}
\def\simlt{\mathrel{\spose{\lower 3pt\hbox{$\mathchar"218$}}
     \raise 2.0pt\hbox{$\mathchar"13C$}}}
\def\simgt{\mathrel{\spose{\lower 3pt\hbox{$\mathchar"218$}}
     \raise 2.0pt\hbox{$\mathchar"13E$}}}
\def\simpropto{\mathrel{\spose{\lower 3pt\hbox{$\mathchar"218$}}
     \raise 2.0pt\hbox{$\propto$}}}

\def\ed{\end{document}}

\def\fsky{f_{\rm sky}}
\def\fcover{f^{\rm cover}}

\def\Ol{\Omega_\Lambda}
\def\Om{\Omega_m}

\def\beq#1{\begin{equation}\label{#1}}
\def\eeq{\end{equation}}
\def\beqa#1{\begin{eqnarray}\label{#1}}
\def\eeqa{\end{eqnarray}}
\def\eq#1{equation~(\ref{#1})}
\def\Eq#1{Equation~(\ref{#1})}
\def\eqn#1{~(\ref{#1})}

\def\d{{\bf d}}

\def\k{{\bf k}}
\def\khat{\widehat{\bf k}}
\def\kperp{{k_\perp}}
\def\l{\ell}

\def\q{{\bf q}}
\def\r{{\bf r}} 
\def\rhat{\widehat{\bf r}}
\def\rperp{{\bf r_\perp}}
\def\s{{\bf s}}
\def\shat{\widehat{\bf s}}
\def\Shat{\widehat{\bf S}}

\def\vv{{\bf v}}
\def\x{{\bf x}}

\def\zhat{\widehat{\bf z}}

\def\B{{\bf B}}

\def\Cnoise{C_0^{\rm noise}}
\def\Clnoise{C_\l^{\rm noise}}

\def\Dmin{D_{\rm min}}
\def\Dmax{D_{\rm max}}

\def\Pnoise{P^{\rm noise}}

\def\S{{\bf S}}
\def\vsigma{{\mathbf\sigma}}

\def\Tsys{T_{\rm sys}}

\def\Wcaret{\hbox{\v{W}}}

\def\Vpix{V_{\rm pix}}
\def\Omegamap{\Omega_{\rm map}}
\def\Omegapix{\Omega_{\rm pix}}

\def\tr{\hbox{tr}\,}

\documentclass[twocolumn,amsmath,nofootinbib]{revtex4} 
\usepackage{amsfonts,amsbsy,epsf} 
\begin{document}






\date{Submitted to Phys.~Rev.~D. June 2 2008, revised March 27 2009, accepted March 31 2009}

\title{The Fast Fourier Transform Telescope}

\author{Max Tegmark}

\address{Dept.~of Physics \& MIT Kavli Institute, Massachusetts Institute of Technology, Cambridge, MA 02139}

\author{Matias Zaldarriaga}

\address{Center for Astrophysics, Harvard University, Cambridge, MA 02138, USA}

\begin{abstract}
We propose an all-digital telescope for 21 cm tomography,
which combines key advantages of both single dishes and interferometers.
The electric field is digitized by antennas on a rectangular grid, after which a series of Fast Fourier Transforms
recovers simultaneous multifrequency images of up to half the sky.
Thanks to Moore's law, the bandwidth up to which this is feasible has now reached about 1 GHz, and will likely continue doubling 
every couple of years.
The main advantages over a single dish telescope are cost and orders of magnitude larger field-of-view, translating into dramatically 
better sensitivity for large-area surveys.
The key advantages over traditional interferometers are cost (the correlator computational cost for an
$N$-element array scales as $N\log_2 N$ rather than $N^2$) and a compact synthesized beam.
We argue that  21 cm tomography could be an ideal first application of a very large Fast Fourier Transform Telescope, which 
would provide both massive sensitivity improvements per dollar and mitigate the off-beam point source foreground problem with its 
clean beam. Another potentially interesting application is cosmic microwave background polarization.
\end{abstract}


  
\maketitle



\setcounter{footnote}{0}

\def\thetamin{\theta_{\rm min}}
\def\thetamax{\theta_{\rm max}}

\section{Introduction}
\label{IntroSec}

Since Galileo first pointed his telescope skyward, design innovations have improved attainable sensitivity, 
resolution and wavelength coverage by many orders of magnitude.
Yet we are still far from the ultimate telescope that simultaneously observes light of all wavelengths 
from all directions, so there is still room for improvement.

From a mathematical point of view, telescopes are Fourier transformers.
We want to know individual Fourier modes $\k$ of the electromagnetic field, as their direction 
$\khat$ encodes our image and their magnitude $k=\omega/c=2\pi/\lambda$ encodes the wavelength,
but the field at a given spacetime point $(\r,t)$ tells us only a sum of all these Fourier 
modes weighted by phase factors $e^{i[\k\cdot\r+\omega t]}$.

Traditional 
telescopes 
perform the spatial Fourier transform from $\r$-space to $\k$-space by approximate analog means
using lenses or mirrors, which are accurate across a relatively small field of view,
and perform the temporal Fourier transform from $t$ to $\omega$ using slits, gratings or 
band-pass filters.
Traditional interferometers used analog means to separate frequencies and measure electromagnetic field correlations 
between different receivers, then Fourier-transformed to $\r$-space digitally, using computers.
In the  tradeoff between resolution, sensitivity and cost, single dish telescopes and interferometers are highly 
complementary, and which is best depends on the science goal at hand.

Thanks to Moore's law, it has very recently become possible to build all-digital interferometers up to about 1 GHz,
where the analog signal is digitized right at each antenna and subsequent correlations and Fourier transforms
are done by computers. In addition to reducing various systematic errors,
this digital revolution enables the ``Fast Fourier Transform Telescope'' or ``omniscope'' that we describe in this paper.
We will show that it acts much like a single dish telescope with a dramatically larger field of view,
yet is potentially much cheaper than a standard interferometer with comparable area.
If a modern all-digital interferometer such as the MWA \cite{MWA} is scaled up to a very large number of antennas $N$,
its price becomes completely dominated by the computing hardware cost for performing of order $N^2$ correlations 
between all its antenna pairs.
The key idea behind the FFT Telescope is that, if the antennas are arranged on a rectangular grid,
this cost can be cut to scale merely as $N\log_2 N$ using Fast Fourier Transforms.
As we will see, this design also eliminates the need for individual antennas that are pointable 
(mechanically or electronically), and has the potential to dramatically improve the 
sensitivity for some applications of future telescopes
like the square kilometer array without increasing their cost.

This basic idea is rather obvious, so when we had it, we wondered why nothing like the massive 
all-sky low-frequency telescope that we are proposing had ever been built.
We have since found other applications of the idea in the astronomy and engineering literature dating as far back as the 
early days of radio astronomy 
\cite{Butler61,May84,Nakajima92,Tanaka00,Nakajima93,Pen04,Takeuchi05,Peterson06,Chang07,SKAprivComm}, 
and it is clear that the answer lies in lack of both computer power and good science applications.
Moore's law has only recently enabled A/D conversion up to the GHz range,
so in older work, Fourier transforms were done by analog means and usually in only one dimension 
(\eg, using a so-called Butler matrix \cite{Butler61}),
%
%
%
severely limiting the number of antennas that could be used.
For example, the 45 MHz interferometer in \cite{May84} used six elements. 
Moreover, to keep the number of elements modest while maintaining large collecting area, the elements themselves would be dishes or 
interconnected antennas that observed only a small fraction of the sky at any one time.
A Japanese group worked on an analog $8\times 8$ FFT Telescope about 15 years ago for studying transient radio sources
\cite{Nakajima92,Nakajima93}, and then upgraded it to digital signal processing aiming for a $16\times 16$ array with a 
field of view just under  $1^\circ$. 
Electronics from this effort is also used in the 1-dimensional 8-element Nasu Interferometer 
\cite{Takeuchi05}.


Most traditional radio astronomy applications involve mapping objects subtending a small angle surrounded by darker background sky,
requiring only enough sensitivity to detect the object itself.
For most such cases, conventional radio dishes and interferometers work well, and an FFT Telescope (hereafter FFTT) is neither necessary nor advantageous.
For the emerging field of 21 cm tomography, which holds the potential to one day overtake the microwave background as 
our most sensitive cosmological probe \cite{Barkana01,ZaldaFurlanettoHernquist03,FurlanettoReview,Loeb06,Santos07,Pritchard08,21cmpars}, the challenge is completely different:
it involves mapping a faint and diffuse cosmic signal that covers all of the sky and needs to be separated from
foreground contamination that is many orders of magnitude brighter, requiring extreme sensitivity and beam control.
This 21cm science application and the major efforts devoted to it by 
experiments such as MWA \cite{MWA}, LOFAR \cite{LOFAR}, PAPER\cite{PAPER}, 21CMA \cite{21CMA}, GMRT \cite{GMRT,GMRT2} and SKA \cite{SKA} makes our paper timely.

An interesting recent development is a North American effort \cite{Peterson06,Chang07} to do 21 cm cosmology with 
a one-dimensional array of cylindrical telescopes that can be analyzed with FFT's, 
in the spirit of the Cambridge 1.7m instrument from 1957,
exploiting Earth rotation to fill in the missing two-dimensional information \cite{Peterson06,Chang07}. 
We will provide a detailed analysis of this design below, arguing that is is complementary to the 2D FFTT at higher frequencies while
a 2D FFTT provides sharper cosmological constraints at low frequencies.


The rest of this paper is organized as follows.
In \Sec{DesignSec}, we describe our proposed design for FFT Telescopes.
In \Sec{ScopeComparisonSec}, we compare the figures of merit of different types of telescopes, and 
argue that the FFT Telescope is complementary to both single dish telescopes and standard interferometers.
We identify the regimes where each of the three is preferable to the other two.
In \Sec{21cmSec}, we focus on the regime where the FFT Telescope is ideal, which 
is when you have strong needs for sensitivity and beam cleanliness but not resolution, 
and argue that
21 cm tomography may be a promising first application for it.
We also comment briefly on cosmic microwave background applications.
We summarize our conclusions in \Sec{ConcSec} and relegate 
various technical details to a series of appendices.

\section{How the FFT Telescope works}
\label{DesignSec}

In this section, we describe the basic design 
and data processing algorithm for the FFT Telescope.
We first summarize the relevant mathematical formalism, then discuss data processing, and conclude by discussing some practical issues.
For a comprehensive discussion of radio interferometry techniques, see e.g. \cite{ThompsonBook}.

\subsection{Interferometry without the flat sky approximation}
\label{FullSkyFormalismSec}

Since the FFT Telescope images half the sky at once, the flat-sky approximation that is common in radio astronomy is not valid.
We therefore start by briefly summarizing the general curved-sky results formalism.
Suppose we have a set of antennas at positions $\r_n$ with sky responses $\B_n(\khat)$ at a fixed 
frequency $\omega=ck$, $n=1,...$, and a sky signal 
$\s(\khat)$ from the direction given by the unit vector $-\khat$ 
(this radiation thus travels in the direction $+\khat$).
The data measured by each antenna in response to a sky signal $\s(\khat)$ is then
\beq{ResponseEq1}
\d_n = \int e^{-i[\k\cdot\r_n+\omega t]}\B_n(\khat)\s(\khat) d\Omega_k.
\eeq
Details related to polarization are covered below in Appendix~\ref{PolarizationSec}, but are irrelevant for 
the present section. For now, all that matters is that $\s(\khat)$ specifies the sky signal, 
$\d_n$ specifies the data that is recorded, and $\B_n(\khat)$ specifies the relation between 
the two.\footnote{If one wishes to observe the sky at frequencies $\nu$
higher than current technology can sample directly ($\simgt 1$ GHz), then
one can extract a bandwidth $\Delta\nu\simlt 1$ GHz in this high frequency range using standard radio engineering techniques
(first an analog frequency mixer multiplies the input signal with that from a local oscillator, then an analog 
low-pass-filter removes frequencies above $\Delta\nu$, and finally the signal is A/D converted).
The net effect of this is simply to replace $e^{-i\omega t}$ in \eq{ResponseEq1}
by $e^{-i(\omega-\omega_0)t}$ for some conveniently chosen local oscillator frequency $\omega_0=2\pi\nu_0$.
It is thus the bandwidth $\Delta\nu$ rather than the actual frequencies $\nu$ that are limited by 
Moore's Law.}
Specifically, $\s$ is the so-called Jones vector (a 2-dimensional complex vector field
giving the electric field components -- with phase -- in two orthogonal directions),
$\d_n$ is a vector containing the two complex numbers measured by the antenna, and
$\B_n(\rhat)$, the so-called {\it primary beam}, is a $2\times 2$ complex matrix field that defines both the polarization response and the 
sky response (beam pattern) of the antenna.
The only properties of \eq{ResponseEq1} that matter for our derivation below are that it is a
linear relation (which comes from the linearity of Maxwell's equations) 
and that it contains the phase factor $e^{-i\k\cdot\r_n}$ (which comes from
the extra path length $\khat\cdot\r_n$ that a wave must travel to get to the antenna location $\r_n$).

The sky signal $\s(\khat)$ has a slow time dependence because the sky 
rotates overhead, because of variable astronomical sources, and because of 
distorting atmospheric/ionospheric fluctuations. However, since these changes
are many orders of magnitude slower than the electric field 
fluctuation timescale $\omega^{-1}$, we can to an excellent approximation treat
\eq{ResponseEq1} as exact for a snapshot of the sky. Below we derive how to recover the snapshot
sky image from these raw measurements; only when coadding different snapshots does one need 
to take sky rotation and other variability into account.

The statements above hold for {\it any} telescope array. 
For the special case of the FFT Telescope, all antennas have approximately identical
beam patterns $(\B_n=\B)$ and lie in a plane, 
which we can without loss of generality take to be the $z=0$ plane so that
$\zhat\cdot\r_n=0$.
Using the fact that 
\beq{dOmegaEq}
d\Omega_k = \sin\theta d\theta d\phi = \frac{dk_x dk_y}{k\sqrt{k^2-\kperp^2}},
\eeq
where $\kperp\equiv\sqrt{k_x^2+k_y^2}$ is the length of the component of the $\k$-vector  
perpendicular to the $z$-axis,
we can rewrite \eq{ResponseEq1} as a 2-dimensional Fourier transform 
\beq{ResponseEq2}
\d_n = \int e^{-i[\q\cdot\x_n+\omega t]}  \frac{\B(\q)\s(\q)}{k\sqrt{k^2-q^2}}  d^2 q = \shat_\B(\x_n)e^{-i\omega t} ,
\eeq
where we have defined the 2-dimensional vectors
\beq{uDefEq}
\q = \left({k_x\atop k_y}\right), \quad \x = \left({x\atop y}\right),
\eeq
and the function 
\beq{sBdefEq}
\s_\B(\q)\equiv \frac{\B(\q)\s(\q)}{k\sqrt{k^2-q^2}}.
\eeq
Here the 2-dimensional function $\s(\q)$ is defined to equal $\s(q_x,q_y,-[k^2-q_x^2-q_y^2]^{1/2})$ 
when $q\equiv|\q|<k$, zero otherwise, and $\B(\q)$ is defined analogously.
$\s_\B$ can therefore be thought of as the windowed, weighted and zero-padded sky signal.
\Eq{ResponseEq2} holds under the assumption that $\B(\khat)$ vanishes for $k_z>0$,  \ie, 
that a ground screen eliminates all response to radiation heading up from below the horizon, 
so that we can limit the integration over solid angle to radiation pointed towards the lower hemisphere.
Note that for our application, the simple Fourier relation of \eq{ResponseEq2} is exact, and that none of 
the approximations that are commonly used in radio astronomy for the so-called ``$w$-term'' 
(see Equation 3.7 in \cite{ThompsonBook}) are needed.

One usually models the fields arriving from different directions as uncorrelated, so that
\beq{SdefEq}
\expec{\s(\khat)\s(\khat')^\dagger} = 
\delta(\khat,\khat')\S(\khat),
\eeq
where $\S(\khat)$ is the $2\times 2$ sky intensity Stokes matrix and the spherical $\delta$-function satisfies 
\beq{SpericalDeltaEq}
\delta(\khat,\khat') = \delta(\q-\q')k\sqrt{k^2-\kperp^2}
\eeq
so that $\int\delta(\khat,\khat') g(\khat')d\Omega'_k = g(\khat)$ for any function $g$. 
Combining \eq{ResponseEq2} with \eq{SdefEq} implies that the correlation between two measurements, 
traditionally referred to as a {\it visibility}, has the expectation value
\beqa{VisibilityEq1}
\expec{\d_m\d_n^\dagger}&=&\int e^{-i\q\cdot(\x_m-\x_n)} 
\frac{\B(\q)^\dagger\S(\q)\B(\q)}{k\sqrt{k^2-q^2}}d^2 q\nonumber\\
&=&\Shat_B(\x_m-\x_n),
\eeqa
where $\Shat_B(\x)$ is the Fourier transform of:
\beq{SBdefEq}
\S_\B(\q)\equiv \frac{\B(\q)^\dagger\S(\q)\B(\q)}{k\sqrt{k^2-q^2}}.
\eeq
is the beam-weighted, projection-weighted and zero-padded sky brightness map.

In summary, things are not significantly more complicated than in 
standard interferometry in small sky patches (where the flat sky approximation is customarily made).
One can therefore follow the usual radio astronomy procedure with minimal
modifications: first measure 
$\Shat_\B(\Delta\x)$ at a large number of baselines $\Delta\x$ corresponding to different antenna
separations $\x_m-\x_n$, then use these measurements to 
estimate the Fourier transform of this function, $\S_\B(\q)$, and
finally recover the desired sky map $\S$ by inverting \eq{SBdefEq}:
\beq{SBInvEq}
\S(\q)= {k\sqrt{k^2-q^2}}{\B(\q)^{-\dagger}\S_\B(\q)\B(\q)^{-1}}.
\eeq

\subsection{FFTT analysis algorithm}

\Eq{VisibilityEq1} shows that the Fourier transformed beam-convolved sky 
$\Shat_B$ is measured at each baseline, \ie, at each separation vector $\x_m-\x_n$ for an antenna
pair. A traditional correlating array with $N_a$ antennas measures 
all $N_a(N_a-1)/2$ such pairwise correlations, and optionally fills in more missing parts of the 
Fourier plane exploiting Earth rotation.
Since the cost of antennas, amplifiers, A/D-converters, {\etc} scales roughly linearly with $N_a$, this means that 
the cost of a truly massive array (like what may be needed for precision cosmology with 21cm tomography \cite{21cmpars})
will be dominated by the cost of the computing power for calculating the correlations, which scales like $N_a^2$.

\begin{figure} 
\centerline{\epsfxsize=8cm\epsffile{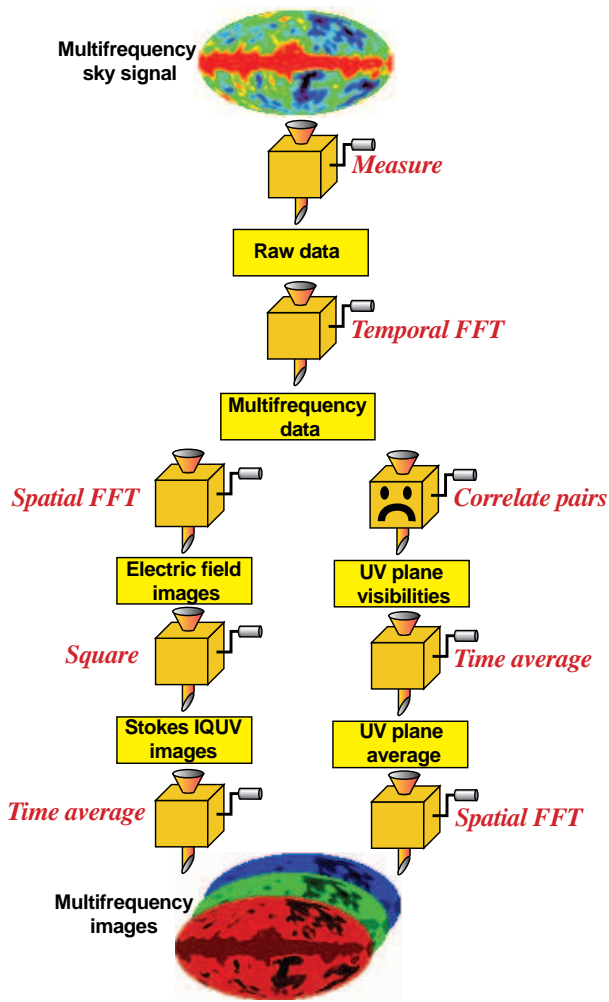}}
\caption[1]{\label{PipelineFig}\footnotesize%
When the antennas are arranged in a rectangular grid as in the FFT Telescope, 
the signal processing pipeline can be dramatically accelerated by eliminating the
correlation step (indicated by a sad face): its computational cost scales as $N_a^2$, because
it must be performed for all {\it pairs} of antennas, whereas all other steps shown scale linearly
with $N_a$. 
The left and right branches recover the same images on average, but with slightly different 
noise. Alternatively, if desired, the FFT Telescope can produce images that are  
mathematically identical to those of the right branch (while retaining the speed advantage) 
by replacing the correlation step marked by the sad face by a spatial FFT, ``squaring,'' and an inverse
spatial FFT.
}
\end{figure}

For the FFT Telescope, the $N_a$ antenna positions $\r_n$ are chosen to form a rectangular grid.
This means that the all $N_a(N_a-1)/2\sim N_a^2$ baselines also fall on a rectangular grid, typically with any given 
baseline being measured by many different antenna pairs.

The sums of $\d_m\d_n^\dagger$ for each baseline can be computed
with only of order $N_a\log_2 N_a$ (as opposed to $N_a^2$) operations by using Fast Fourier Transforms.
Essentially, what we wish to measure in the Fourier plane are the 
antenna measurements (laid out on a 2D grid) convolved with themselves, 
and this naively $N_a^2$ convolution can be 
reduced to an FFT, a squaring, and an inverse FFT.

In fact, \eq{ResponseEq2} shows that 
after FFT-ing the 2D antenna grid of data $\d_n$, 
one already has the two electric field components $\s_\B(\q)$ from each sky direction,  
and can multiply them to measure the sky intensity from each direction (Stokes $I$, $Q$, $U$ and $V$)
without any need to return to Fourier space, as illustrated in \fig{PipelineFig}.
This procedure is then repeated for each time sample and each frequency, 
and the many intensity maps at each frequency are 
averaged (after compensating for sky rotation, ionospheric motion, \etc) to improve signal-to-noise.

It should be noted that the computational cost for the entire FFT Telescope signal processing pipeline is 
(up to some relatively unimportant log factors) merely proportional to the total number of numbers measured by 
all antennas throughout the duration of the
observations. In particular, the time required for the spatial FFT operations is of the same order as the time required for the
time-domain FFT's that are used to separate out the different frequencies from the time signal using 
standard digital filtering.
If the antennas form an $n_x\times n_y$ rectangular array, so that $N_a=n_x n_y$, and each antenna measures $n_t$ different time samples 
(for a particular polarization), then it is helpful to imagine this data arranged in a 3-dimensional $n_x\times n_y\times n_t$ block.
The temporal and spatial FFT's (left branch in \fig{PipelineFig}) together correspond to a 3D FFT of this block, 
performed by three 1-dimensional FFT operations:
\begin{enumerate}
\item For each antenna, FFT in the $t$-direction.
\item For each time and antenna row, FFT in the $x$-direction.
\item For each time and antenna column, FFT in the $y$-direction.
\end{enumerate}
One processes one such block for each of the two polarizations.
These three steps each involve of order $n_t n_x n_y$ multiplications (up to order-of-unity factors $\log n_t$, $\log n_x$ and $\log n_y$),
and it is easy to show that the number of operations for the three steps combined scales as $(n_t n_x n_y)\log(n_t n_x n_y)$, \ie, depends only on the
total amount of data $n_t n_x n_y$.
After step 3, one has the two electric field components from each direction at each frequency.
Phase and amplitude calibration of each antenna/amplifier system is normally performed after step 1.
If one is interested in sharp pulses that are not well-localized in frequency, one may opt
to skip step 1 or perform a broad band-pass filtering rather than a full spectral separation.

The FFT Telescope cuts down not only on CPU time, but also on data storage costs,
since the amount of data obtained at each snapshot scales as number of time samples taken times 
$N_a$ rather than $N_a^2$.

In a conventional interferometer, antennas are correlated only with other antennas and not with themselves, to eliminate noise bias.
This can be trivially incorporated in the FFTT analysis pipeline as well by setting the pixel at the origin of the UV plane (corresponding to zero baseline)
to zero, and is mathematically equivalent to removing the mean form the recovered sky map.


\subsection{Practical considerations}

Although we have laid out the mathematical and computational framework for an FFT Telescope above, 
there are a number of practical issues that require better understanding before building a massive scale FFT Telescope.

As we will quantify in \Sec{ScopeComparisonSec} below, the main advantages of an FFT Telescope relative to single dish telescopes
and conventional interferometers emerge when the number of antennas $N_a$ is very large.
A successful FFTT design should therefore emphasize simplicity and mass-production, 
and minimize hardware costs.
To exploit the FFT data processing speedup, care must be taken to make the antenna array as uniform as possible.
The locations $\r_i$ of the antennas need to be kept in a planar rectangular
grid to within a small fraction of a wavelength, so when selecting the construction site, it is important that the land is 
quite flat to start with, that bulldozing is feasible, and
that there are no immovable obstacles.
It is equally important that the sky response $\B(\r)$ be close to identical for all antennas.
A ground screen, which can simply consist of cheap wire mesh laid out flat under the entire array, should therefore extend sufficiently
far beyond the edges of the array that it can to reasonable accuracy be modeled as an infinite reflecting plane, 
affecting all antennas in the same way.
The sky response $\B(\r)$ of an antenna will also be affected by the presence of neighbors: whereas 
the response of antennas in the central parts of a large array will be essentially identical to one another (and 
essentially identical to that for an antenna in the middle of an infinite array), antennas near the edges of the array 
will have significantly different response.
Instead of complicating the analysis to incorporate this, it is probably more cost effective to surround the desired array
with enough rows of dummy antennas that the active ones can be accurately modeled as being in an infinite array.
These dummy antennas could be relatively cheap, as they need not be equipped with amplifiers or other electronics 
(merely with an equivalent impedance), and no signals are extracted from them.

The FFT algorithm naturally lends itself to a a rectangular array of antennas. However, this rectangle need not be square; 
we saw above that the processing time is independent of the shape of the rectangle, depending only on the total number of antennas,
and below we will even discuss the extreme limit where the telescope is one-dimensional.
Another interesting alternative to a square FFTT telescope is a circular one, 
consisting of only those $\pi/4\approx 79\%$ of the antennas in the square grid that lie within a circle inscribed in the square.
This in no way complicates the analysis algorithm, as the FFT's need to be zero-padded in any case, and increases the computational cost for a given 
collecting area by only about a quarter. The main advantage is a simple rotationally invariant synthesized beam as discussed below.
Antennas can also be weighted in software before the spatial FFT do create beams with other desired properties; for example, edge tapering can be used to make the
beam even more compact. 
A third variant is to place the antennas further apart to gain resolution at the price of undersampling the Fourier plane and picking up sidelobes.

\begin{table*}
\bigskip
\noindent
{\bf Table~\ScopeComparisonTable} -- How telescope properties scale with dish size $D$, collecting area $A$ and wavelength $\lambda$.
We assume that the standard interferometer has $N_a\sim A/D^2$ separate dishes with a maximum separation $\Dmax$ that together cover a fraction
$\fcover\sim A/\Dmax^2$ of the total array region rather uniformly.  
\begin{center}   
{\footnotesize
\begin{tabular}{|ll|c|c|c|c|}
\hline
			&			&\multicolumn{2}{c|}{Single Dish Telescopes}		&\multicolumn{2}{c|}{Interferometers}\\
\hline
			&			&Single			&Maximal				&						&Standard\\
			&			&Receiver		&Focal Plane				&FFT						&Interferometric\\
			&			&Telescope		&Telescope				&Telescope					&Telescope\\
\hline
Resolution		&$\thetamin$		&$\frac{\lambda}{D}$	&$\frac{\lambda}{D}$			&${\frac{\lambda}{D}}$				&$\frac{\lambda}{\Dmax}$\\
Field of view		&$\thetamax$		&$\frac{\lambda}{D}$	&$\left(\frac{\lambda}{D}\right)^{1/3}$	&$1$						&$\frac{\lambda}{D}$\\
Resolution elements	&$n$			&$1$			&$\left(\frac{D}{\lambda}\right)^{4/3}$	&$\left(\frac{D}{\lambda}\right)^2$		&$\left(\frac{\Dmax}{D}\right)^2$\\
Etendu			&$A\Omega$		&$\lambda^2$		&$D^{4/3}\lambda^{2/3}$			&$D^2$						&$\frac{\lambda^2A}{D^2}$\\
Sensitivity		&$\Cnoise$		&$\lambda\Tsys^2$	&$\frac{\lambda^{7/3}\Tsys^2}{A^{2/3}}$	&$\frac{\lambda^3\Tsys^2}{A}$			&$\frac{\lambda D^2\Tsys^2}{A\fcover}$\\
Cost			&$\$$			&$A^{1.35}$		&$A^{1.35}$				&$A$						&$A^2$\\
\hline
\end{tabular}
}
\end{center}     
\end{table*}

\begin{figure}[ht]
\centerline{\epsfxsize=\figsize\epsffile{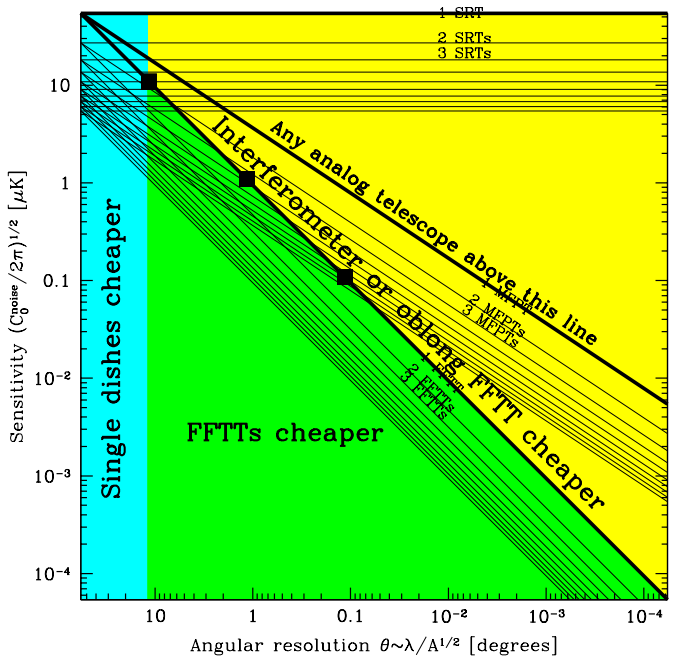}}
\caption{\label{ScopeComparisonFig}
Angular resolution and sensitivity are compared for different telescope designs, assuming 
that half the sky is surveyed during 4000 hours at an observing frequency $150$ MHz, with
0.1\% bandwidth and 200K system temperature. 
Since $\delta T_\l=\sqrt{\l^2\Cnoise/2\pi}$, $\delta T_\l=10$mK at $\l=1000$ corresponds
to $10\mu K$ on the vertical axis.
The parameters of any analog telescope (using focusing optics rather than digital beamforming) lie
in the upper right triangle between the limiting cases of the single receiver telescope 
(SRT; heavy horizontal line) and the single dish telescope with a maximal focal plane 
(MFPT; heavy line of slope 2/3). The parameters of a 
Fast Fourier transform telescope (FFTT) lie on the heavy horizontal line of slope -1, with solid squares
corresponding to squares FFTTs of side 10m, 100m and 1000m, respectively. Moving their antennas further apart
(reducing $\fcover$ with $A$ fixed) would move these squares along a $45^\circ$ line up to the right.
Improved sensitivity at fixed resolution can be attained by building multiple  
telescopes (thin parallel lines correspond to 2, 3,...,10 copies).
As explained in the text, SDTs, SITs and FFTTs are complementary:
the cheapest solution is offered by SDTs for low resolution, 
FFTTs for high sensitivity $(\Cnoise)^{1/2}\simlt\theta\times 2\mu$K, 
and elongated FFTTs or standard interferometers for high resolution 
$\theta\simlt (\Cnoise)^{1/2}/2\mu$K.
}
\end{figure}

\section{Comparison of different types of telescopes}
\label{ScopeComparisonSec}

\subsection{Telescopes generalized}

In this section, we compare the figures of merit (resolution, sensitivity, cost, \etc)
of different types of telescopes, summarized in Table~\ScopeComparisonTable, and 
argue that the FFT Telescope is complementary to both single dish telescopes and standard interferometers.
We identify the regimes where each of the three is preferable to the other two, as summarized in
\fig{ScopeComparisonFig}.

It is well-known that all telescopes can be analyzed within a single unified formalism
that characterizes their linear response to sky signals and their noise properties.
In particular, a single dish telescope can be thought of as an interferometer, where every little piece of the collecting area
is an independent antenna, and the correlation is performed by approximate analog means using curved mirrors.
This eliminates the costly computational step, but the approximations involved are only valid in a limited field of view 
(Table~\ScopeComparisonTable). Traditional interferometers can attain larger field of view and better resolution for a given collecting area,
but at a computational cost. The FFT Telescope is a hybrid of the two in the sense that it combines the resolution of
a single dish telescope with the all-sky field-of-view of a dipole interferometer --- at a potentially much lower cost than either a single dish or
a traditional interferometer of the same collecting area.
Let us now quantify these statements, starting with angular resolution and its generalization and then turning to sensitivity and cost.
We first briefly review some well-known radio astronomy formalism that is required for our applications.

%

\subsection{Angular resolution and the beam function $B_\l$}

\begin{figure} 
\centerline{\epsfxsize=\figsize\epsffile{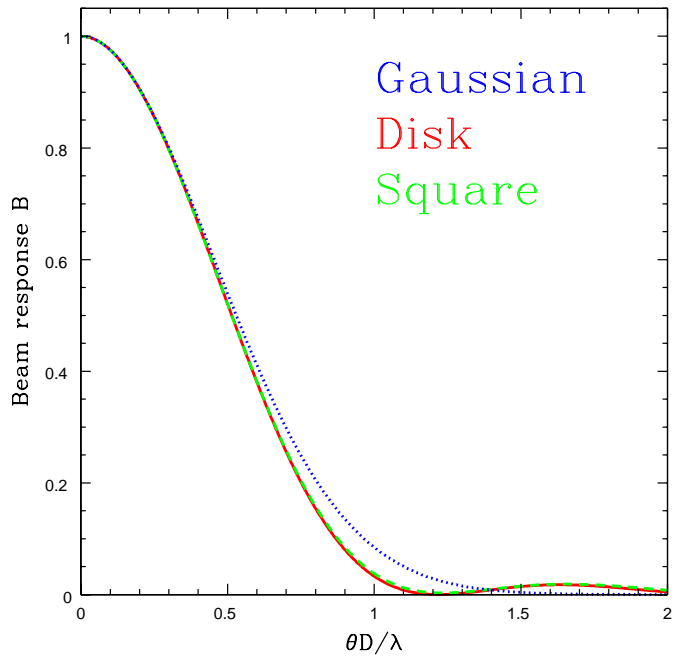}}
\caption[1]{\label{AiryFig}\footnotesize%
The familiar Airy pattern that constitutes the sky response of a circular telescope dish
of diameter $D$
is compared with the azimuthally averaged response of a square telescope and one with a Gaussian tapered aperture.
The square has side $0.87D$ to have the same FWHM, and the Gaussian has standard deviation $0.45D$ to give
comparable response for $\theta D/\lambda\ll 1$.
}
\end{figure}

\begin{figure} 
\centerline{\epsfxsize=\figsize\epsffile{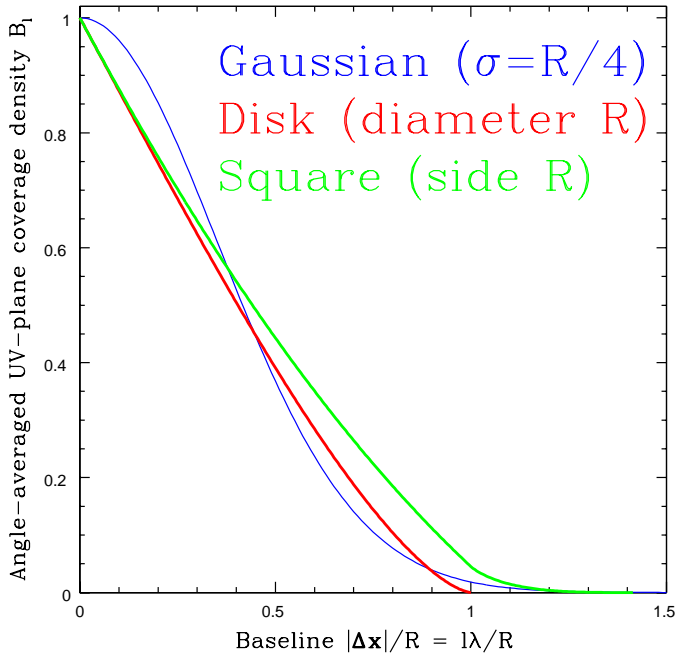}}
\caption[1]{\label{uvCoverageFig}\footnotesize%
The angle-averaged UV plane sensitivity $f_\l$ (the relative number of baselines width different lengths) 
is compared for telescopes of different shapes. 
}
\end{figure}

The angular resolution of the telescopes we will compare are all much better than a radian, so
we can approximate the sky as flat for the purposes of this section.
If we ignore polarization, then it is well-known that the response of an interferometer to radiation intensity coming from near the local
zenith\footnote{If we are imaging objects much smaller than a radian
centered at a zenith angle $\theta$, we recover the same formula as above, but with the synthesized beam
compressed by a factor $\cos\theta$ in one direction, as the source effectively sees the array at a slanting angle, compressed 
by $\cos\theta$ in one direction.
} and
traveling in the direction $\k$ is $\Wcaret(k_x,k_y)$, the inverse Fourier transform of the function $W(\Delta x)$ 
that gives the distribution of baselines $\Delta\x$.
%
For the classic example of a single dish telescope of radius $R$, 
this formula gives
\beq{SDTbeamEq} 
\Wcaret(k_x,k_y)=\left[\frac{2J_1(Rk_\perp)}{Rk_\perp}\right]^2,
\eeq
the famous Airy pattern plotted in \fig{AiryFig}.
Here $\kperp=(k_x^2+k_y^2)^{1/2}=2\pi\theta/\lambda$, where $\theta$ is the angle to the zenith.
When the beam is asymmetric, we will mainly be interested in the azimuthally averaged beam which again depends only on $\theta$;
the result for a square telescope like the fully instrumented FFTT plotted 
for comparison\footnote{
For a telescope with a square dish of side $D=2R$, convolving the square with itself gives the 
baseline distribution 
\beq{FFTfEq1} 
W({\bf\Delta x})\propto(2R-|\Delta x|)(2R-|\Delta y|)
\eeq
when $|\Delta x|<2R$ and $|\Delta y|<2R$, zero otherwise.
Writing ${\bf\Delta x}=r(\cos\varphi,\sin\varphi)$ and averaging over the azimuthal angle $\varphi$ gives
\beq{FFTTfEq2} 
W(r)=\left\{
\begin{tabular}{ll}
$1-\frac{1}{\pi}\left(4-\frac{r}{D}\right)\frac{r}{D}$		&if $r\le D$,\\
$1-\frac{4}{\pi}\left[\frac{1}{2}-\sqrt{\frac{r^2}{D^2}-1}+\frac{r^2}{4D^2}+\cos^{-1}\frac{D}{r}\right]$	 &if $r\ge D$,
\end{tabular}
\right.
\eeq
which is plotted in \fig{uvCoverageFig}.
The synthesized beam is simply
\beq{FFTTwhatEq} 
\Wcaret(\k)=j_0(R k_x)j_0(Rk_y),
\eeq
and the azimuthal average of this function is plotted in \fig{AiryFig}.
}.
The figure also shows a Gaussian beam, which may be a better approximation for an optical telescope when the seeing is poor.

For these three cases, the shapes are seen to be sufficiently similar that, for many purposes, all one needs to know about the beam can be
encoded in a single number specifying its width.
The most popular choices in astronomy are summarized in Table~\ResolutionComparisonTable:
the rms (the root-mean-squared value of $\theta$ averaged across the beam), 
the FWHM (twice the $\theta$-value where $B(\theta)$ has dropped to half its central value)
and the first null (the smallest $\theta$ at which $B(\theta)=0$).
We will mainly focus on the FWHM in our cost comparison below.

\begin{table*}
\bigskip
\noindent
{\bf Table~\ScopeComparisonTable} -- Different measures of angular resolution, measured in units of $D/\lambda$.
\begin{center}   
{\footnotesize
\begin{tabular}{|l|ccc|}
\hline
				&rms		&FWHM		&First null\\
\hline
Disk of diameter $D$		&0.53		&$1.03$		&1.22\\
Square of side $D$		&0.49		&$0.89$		&1.07\\
Gaussian with $\sigma=D$	&1		&$2.35$	        &$\infty$\\
\hline
\end{tabular}
%
}
\end{center}     
\end{table*}

The primary beam $\B(\q)$ that was introduced in \Sec{FullSkyFormalismSec} can itself be
derived from this same formalism by considering each piece of an antenna as an independent
element. For example, a single radio dish has $\B\propto\Wcaret$ with $\Wcaret$ given by \eq{SDTbeamEq}
modulo polarization complications. To properly compute the polarization response that is
encoded in the matrix $\B$, the full 3-dimensional structure of the antenna and how it is
connected to  the two amplifiers must be taken into account, and the presence of nearby
conducting objects affects $\B$ as well.


For applications like CMB and 21cm mapping, where one wishes to measure a cosmological power spectrum, the key aspect of the
synthesized beam that matters is how sensitive it is to different angular scales $\l$ and their associated spherical harmonic
coefficients. This response to different angular is encoded in the spherical harmonic expansion of the synthesized beam $\Wcaret(k_x,k_y)$.
If the synthesized beam is rotationally symmetric (or made symmetric by averaging observations with different orientations as Earth rotates), 
then its spherical harmonic coefficients $\Wcaret_{\l m}$ vanish except for $m=0$, 
and we only need to keep track of the so-called {\it beam function}, the coefficients $B_\l\equiv\Wcaret_{\l 0}$ plotted in \fig{uvCoverageFig}.
In the flat-sky approximation, this beam function for a rotationally symmetric synthesized beam reduces to the two-dimensional 
Fourier transform of $\Wcaret(k_x,k_y)$, which is simply the baseline distribution $W(\Delta x)$:
\beq{BlApproxEq}
B_\l\approx W\left(\l/k\right).
\eeq
\Fig{uvCoverageFig} shows $B_\l$ for the circular, square and Gaussian aperture cases mentioned above.
WMAP and many other CMB experiments have published detailed measurements of their beam functions $B_\l$ (\eg, \cite{wmap5beams}), many of which are 
fairly well approximated by Gaussians.
For interferometers, the beam functions can be significantly more interesting.
Since 
$B_\l^2$ scales simply as the number of baselines at different separations, more complicated synthesized beams involving more than one scale can be designed if desirable.

\subsection{Sensitivity}

\subsubsection{How the noise power spectrum is defined and normalized}

The sensitivity of an arbitrary telescope to signals on various angular scales is quantified by its noise power spectrum
$\Clnoise$. 
If the telescope were to make a uniformly observed all-sky map, then $\Clnoise$ would be the variance (due to detector noise) 
with which a spherical harmonic coefficient $a_{\l m}$ could be measured. For a map that covers merely a small sky patch, 
the corresponding noise power spectrum is the $\Clnoise$ that would result if the whole sky were observed with this same sensitivity.
Without loss of generality, we can factor the noise power spectrum as \cite{Knox95,wiener}
\beq{ClnoiseEq1}
\Clnoise = \Cnoise B_\l^{-2},
\eeq
where $B_\l$ is the beam function from the previous section, and $\Cnoise$ is an overall normalization constant.
To avoid ambiguity in this factorization, we normalize the beam function $B_\l$ so that its maximum value equals unity.
For a single dishe, the maximum is always at $\l=0$. This gives the normalization 
$B_0=1$, which given \eq{BlApproxEq}, which 
means that the synthesized beam $\Wcaret(\q)$ integrates to unity and that 
we can interpret the signal as measuring a weighted average of the true sky map.
In all cases, our normalization 
Most interferometers have $B_0=0$ and thus no sensitivity to the mean; in many such cases, 
$B_\l$ is roughly constant on angular scales $\l^{-1}$ much larger than the synthesized beam but much smaller than the 
primary beam, taking its maximum on these intermediate scales.

This seemingly annoying lack of sensitivity to the mean is a conscious choice and indeed a key advantage of 
interferometers. The mean sensitivity can optionally be retained by simply including the antenna autocorrelations in the analysis 
(\ie, not explicitly setting the pixel at the origin of the $(u,v)$ plane equal to zero), but this pixel normally contains a large
positive bias due to noise that is difficult to accurately subtract out. 
In contrast, the noise in all other $(u,v)$ pixels normally has zero mean, because the noise in different antennas is uncorrelated.
Since single-dish telescopes cannot exclude this zero mode, they often require other approaches to mitigate this noise bias, such as 
rapid scanning or d beamswitching.

\subsubsection{How it depends on experimental details}

Consider a telescope with total collecting area $A$ observing for a time $\tau$ with a bandwidth $\Delta\nu$ around some frequency $\nu=c/\lambda$.
If this telescope performs a single pointing in some direction, then the noise power spectrum for 
this observed region is \cite{ZaldaFurlanettoHernquist03}:
\beq{CnoiseEq1}
\Cnoise 
=\frac{\lambda^2\gamma^2\Tsys^2}{A\fcover\tau\Delta\nu}.
\eeq
Here $\gamma$ is a dimensionless factor of order unity that depends on the convention used to define the telescope system temperature $\Tsys$; 
below we simply adopt the convention where $\gamma=1$.
For a single-dish telescope and for a maximally compact interferometer like the FFTT, $\fcover=1$.
For an interferometer where the antennas are rather uniformly spread out over a larger circular area, $\fcover$ is the fraction of this
area that they cover; if there are $N_a$ antennas with diameter $D$ in this larger area of diameter $\Dmax$, we thus have 
$\fcover=N_a (D/\Dmax)^2$ and total collecting area $A=N_a\pi(D/2)^2$. 
For a general interferometer the noise power spectrum depends on the distribution of baselines and could be a complicated function of $\l$. We are absorbing all $\l$-dependence into the beam function $B_\l$ as per \eq{ClnoiseEq1}.

If instead of just pointing at a fixed sky patch, the telescope scans the sky (using Earth rotation and/or pointing) 
to map a solid angle $\Omegamap$ that exceeds
its field-of view $\Omega$, and spends roughly the same amount of time covering all parts of the map, then 
a given point in the map is observed a fraction $\Omega/\Omegamap$ of the time.
The resulting noise power spectrum for the map is then
\beq{CnoiseEq2}
\Cnoise 
=\frac{4\pi}{\eta}\frac{\lambda^3\fsky\Tsys^2}{\fcover A\Omega c\tau}.
\eeq
Here $\fsky\equiv\Omegamap/4\pi$ is the fraction of the sky covered by the map, and we have 
introduced the dimensionless parameter $\eta\equiv\Delta\nu/\nu=\Delta\nu\,c/\lambda$ to denote the relative bandwidth.


\subsection{The 3D noise power spectrum $\Pnoise$}

For 21cm applications, it is also important to know the three-dimensional 
noise power spectrum the ``data cube'' mapped by treating the frequency as the radial direction
(the higher the frequency, the larger the redshift and hence the larger the distance to the hydrogen gas responsible
for the 21 cm signal).
In a comoving volume of space subtending a small angle $\theta\ll 1$ and a small 
redshift range $\Delta z/z\ll 1$ centered around $z_*$, we can linearize the relation between 
the comoving coordinate $\r$ and the observed quantities $(\theta_x,\theta_y,\nu)$ (\eg, \cite{21cmpars}):
\beqa{GeometryLinearizationEq}
\Delta\rperp	&=&d_A(z_*)\Delta{\bf\Theta}\\
\Delta r	&=&y(z_*)\Delta\nu.\label{GeometryLinearizationEq2}
\eeqa
Here $\Delta{\bf\Theta}\equiv(\theta_x,\theta_y)=(\khat_x,\khat_y)$ gives the angular distance away from the center of the field being imaged,
and $\Delta\rperp$ is the corresponding comoving distance transverse to the line of sight. 
$d_A(z)$ is the comoving angular diameter distance to redshift $z$, and
\beq{yDefEq} 
y(z) = \frac{\lambda_{21}(1+z)^2}{H(z)},
\eeq
where $\lambda_{21}\approx 21$ cm is the rest-frame wavelength of the 21 cm line, and $H(z)$ is the cosmic expansion rate at
redshift $z$.
In Appendix~\ref{GeometryAppendix}, we show that these two conversion functions can be accurately approximated by
\beqa{GeometryApproxEq}
d_A(z_*)	&\approx&14.8\Gpc - \frac{16.7\Gpc}{(1+z)^{1/2}},\\  
y(z)		&\approx&\frac{\lambda_{21}(1+z)^{1/2}}{\Omega_m^{1/2}H_0}
\approx\frac{18.5\Mpc}{1\,\MHz}\left(\frac{1+z}{10}\right)^{1/2}\label{GeometryApproxEq2}
\eeqa
for the $z\gg 1$ regime most relevant to 21 cm tomography given the 
flat concordance cosmological parameter values $\Omega_m=0.25$ and $H_0=72$ km$\,$s$^{-1}$Mpc$^{-1}$ \cite{lrg,wmap5pars}.


If a 2-dimensional map is subdivided into pixels of area $\Omegapix$ and the noise is uncorrelated with variance $\sigma^2$ in these
pixels, then 
\beq{OmegapixEq}
\Clnoise=\sigma^2\Omegapix
\eeq
for angular scales $\l$ well above the pixel scale.
Analogously, if a 3-dimensional map is subdivided into pixels (voxels) of volume $\Vpix$ and the noise is uncorrelated with variance $\sigma^2$ in them,
then 
\beq{VpixEq}
\Pnoise=\sigma^2\Vpix
\eeq
on length scales well above the pixel scale.
Since the volume of a 3D pixel is $\Vpix=(d_A^2\Omegapix)\times (y\Delta\nu)$, \ie, 
its area times its depth,
combining equations\eqn{CnoiseEq2}, \eqn{OmegapixEq} and\eqn{VpixEq} gives
the large-scale noise power spectrum 
\beq{PnoiseEq}
\Pnoise=\frac{4\pi\fsky\lambda^2\Tsys^2 y\,d_A^2}{A\Omega\fcover\tau}.
\eeq

When 2D and 3D power spectra are discussed in the cosmology literature,
it is popular to introduce corresponding quantities 
\beqa{dTdefEq}
(\delta T_\l)^2	&\equiv&\frac{\l(\l+1)}{2\pi}C_\l,\\
\Delta(k)^2	&\equiv&\frac{4\pi k^3}{(2\pi)^3}P(k)\label{DeltaDefEq},
\eeqa
which give the variance contribution per logarithmic interval in scale.
One typically has $\delta T_\l\sim\Delta(k)$
when both the angular scale $\l$ and the bandwidth $\Delta\nu$ are chosen to match the length scale $\Delta r=2\pi/k$,
\ie, when $\l=k/d_A$ and $\Delta\nu=2\pi/ky$.
Beware that here (and only here) we use $k$ to denote the wavenumber of cosmic fluctuations, while 
everywhere else in this paper, we use it to denote the wave vector of electromagnetic radiation.


\subsubsection{Sensitivity to point sources}

It is obviously good to have a small noise power spectrum $\Clnoise$ and a large field of view.
However, the tradeoff between these two differs depending on the science goal at hand. 
Below we mention two cases of common interest.

If one wishes to measure the flux $\phi$ from an isolated point source, it is easy to show that the attainable accuracy $\Delta\phi$ is
\beq{PSsensitivityEq}
\Delta\phi = \sqrt{4\pi\Cnoise}\left[\sum_{\l=0}^\infty (2\l+1)B_\l^2\right]^{-1/2}.
\eeq
In the approximation of a Gaussian beam $B_\l=e^{-\theta^2\l^2/2}$ with rms width $\theta\ll 1$,
this simplifies to
\beqa{PSsensitivityEq2}
\Delta\phi&\approx&\theta\sqrt{4\pi\Cnoise}=4\pi\Tsys\theta\sqrt{\frac{\lambda^3\fsky}{A\Omega\fcover\eta c\tau}}\nonumber\\
&\sim&\Tsys\sqrt{\frac{\lambda^4\fsky}{A^2\Omega\tau\Delta\nu}}.
\eeqa
In the last step, we used the fact that the angular resolution $\theta\sim\lambda/(A/\fcover)^{1/2}$.
The total information (inverse variance) in the map about the point source flux thus scales as
$A^2\Omega\tau\Delta\nu$. That this information is proportional to the field of view $\Omega$, the 
observing time $\tau$ and te bandwidth $\Delta\nu$ is rather obvious. That it scales like the 
collecting area as $A^2$ rather than $A$ is because every baseline carries an equal amount of 
information about the flux $\phi$, and the number of baselines scales quadratically with the area. 
It is independent of $\fcover$ because it does not matter how long the baselines are; therefore
the result is the same regardless of where the antennas are placed.
This last result also provides intuition for the $\fcover$-factor in \eq{CnoiseEq1}:
since $\theta^2\Clnoise$ is independent of $\fcover$ and 
$\theta\propto\lambda/(A/\fcover)^{1/2}\propto\sqrt{\fcover}$, we must have 
$\Clnoise\propto 1/\fcover$. As $\fcover$ drops and the same total amount of information 
is spread out over an area in the UV plane that is a factor $1/\fcover$ larger, 
the information in any given $\l$-bin that was previously observed must drop by the same factor, 
increasing its variance by a factor $1/\fcover$.




\subsubsection{Power spectrum sensitivity }
\label{DClSec}


For CMB and 21 cm applications, one is interested in measuring the power spectrum $C_\l$ of the sky signal.
The accuracy with which this can be done depends not only on $\Clnoise$, but also on the signal $C_\l$ itself (which contributes sample variance)
and on the mapped sky fraction $\fsky$. The average power spectrum across a band consisting of $\Delta\l$ multipoles centered around $\l$ can 
be measured to precision \cite{Scott94,strategy}
\beq{dClEq}
\Delta C_\l \approx \sqrt{\frac{2}{(2\l+1)\Delta\l\fsky}}\left(C_\l+\Clnoise\right).
\eeq
Since $\Clnoise\propto\fsky$, there is an optimal choice of $\fsky$ that minimizes $\Delta C_\l$. In cases where $\fsky<1$ is optimal, this best choice
corresponds to $\Clnoise\sim C_\l$, so that sample variance and noise make comparable contributions \cite{Scott94,strategy}.
This means that optimized measurements tend to fall into one of three regimes:
\begin{enumerate}
\item No detection: $\Delta C_\l\simgt C_\l$ even when $\fsky$ is made as small as the telescope permits. Upper limit $\propto\Clnoise$.
\item Improvable detection: $\Clnoise\sim C_\l$, and $\Delta C_\l\simpropto\fsky^{-1/2}\simpropto(\Clnoise)^{1/2}$.
\item Cosmic variance limited detection: $\Clnoise\ll C_\l$, and further noise reductions do not help.
\end{enumerate}
The regime depends normally depends on $\l$, since $C_\l$ and $\Clnoise$ tend to have different shapes.
For example, the WMAP measurement of the unpolarized CMB is in regimes 1, 2 and 3 at
$\l\sim1000$, $\l\sim300$ and $\l\sim100$, respectively. 
 




\subsection{Field of view $\Omega$}

The field of view of a telescope is the solid angle $\Omega$ that it can map in a single pointing.
For a telescope with a single dish of diameter $D$ and a single receiver/detector pixel in its focal plane (a dish for satellite TV reception, say), 
the receiver will simply map a sky patch corresponding to the angular resolution $\sim\lambda/D$, giving 
$\Omega\sim(\lambda/D)^2$. 
The opposite extreme is to fill the entire focal plane with receivers, as is often done for, e.g., microwave and optical telescopes.
In Appendix~\ref{FOVappendix}, we show that the largest focal plane possible covers an angle of order $(\lambda/D)^{1/3}$, corresponding to 
$\Omega\sim(\lambda/D)^{2/3}$. 
This upper bound comes from the fact that the analog Fourier transform performed by telescope optics is only approximate.
Many actual multi-receiver telescopes fall somewhere between these two extremes. In summary, 
single-dish telescopes have a field of view somewhere in the range
\beq{OmegaRangeEq}
\left(\frac{\lambda}{D}\right)^2\simlt\Omega\simlt\left(\frac{\lambda}{D}\right)^{2/3}.
\eeq
We refer to the two extreme cases in this inequality as the single receiver telescope (SRT) and the maximal focal plane telescope (MFPT), respectively.

Since the performs its Fourier transform with no approximations, it can in principle observe the entire sky above the horizon, corresponding to 
$\Omega=2\pi$. However, the useful field of view is only of order half of this, because the image quality degrades near the horizon:
viewed from a zenith angle $\theta$, one dimension of the telescope appears foreshortened by a factor $\cos\theta$, causing loss of both 
angular resolution and collecting area (and thus sensitivity) near the horizon. 

\subsection{Cost}
\label{CostSec}

Detailed cost estimates for telescopes are notoriously difficult to make, and will not be attempted 
here. We will instead limit our analysis to the approximate scaling of cost with collecting 
area, as summarized in Table~\ScopeComparisonTable, 
which qualitatively determines which telescopes are cheapest in the different parts of the
parameter space of \fig{ScopeComparisonFig}.

For a single-dish telescope, the cost usually grows slightly faster than linearly with area.
Specifically, it has been estimated that the cost $\simpropto A^{1.35}$ for radio telescopes \cite{Wilkinson90}.

For a standard interferometric telescope consisting of $N$ separate dishes, the total 
cost for the dishes themselves is of course proportional to $N$.
However, the cost for the correlator hardware that computes the correlations between all the 
$N(N-1)/2$ pairs of dishes scales as $N^2$, and thus completely dominates the total cost in the large $N$ limit
that is the focus of the present paper
(already at the modest scale of the MWA experiment, where $N=512$, the $N$ and $N^2$ parts of the hardware 
cost are comparable).
For fixed dish size, the total collecting area $A\propto N$ so that 
the cost $\propto A^2$.
f
For an FFT Telescope, the cost of antennas, ground screen and amplifiers are all proportional to the number of 
antennas and hence to the area. As described in \Sec{DesignSec}, the  computational hardware 
is also proportional to the area, up to some small logarithmic factors that we to first approximation can ignore.

The above-mentioned approximate scalings are of course only valid over a certain range. 
All telescopes must have $A\simgt\lambda^2$. The cost of single dishes grows more rapidly once their structural integrity
becomes an issue --- for example, engineering challenges appear to make 
a single-dish radio telescope with $A=1\>$km$^2$ daunting with current technology \footnote{However, an interesting design for which this might be
feasible has been proposed in \cite{Legg98}, where an almost flat telescope rests close to the ground and the focal plane is carried by a steerable
Helium balloon.}, 
and for an FFTT with diameter 
$A^{1/2}\gg 10\>$km, compensating for Earth's curvature could become 
a major cost.\footnote{If Earth were a perfect sphere of radius $R\approx 6400\>$km, then a planar telescope of 
radius $r$ would be a height 
\beq{HeightOverGroundEq}
h\approx\frac{r^2}{2R}\approx 8\>{\rm m}\left(\frac{r}{10\>{\rm km}}\right)^2
\eeq
above the ground at its edges. 
If the telescope is not planar, one cannot use the straightforward FFT analysis method.
In practice, this might only be a problem if {\it both} of the dimensions of the FFTT are $\gg 10\>$km:
as long as the telescope can be kept flat in the {\it narrowest} dimension, it will have no 
intrinsic (Gaussian) curvature even if the the telescope has Earth's circular shape in its wide direction.
An interesting question for future work is whether some algorithm incorporating an FFT along the long axis
can be found that provides and exact and efficient recovery of the sky map for this generalized case.
}
Finally, an all-digital telescope like the FFTT is currently limited to by Moore's law for computer processing speed to 
frequencies below a few GHz, and analog interferometry has not yet been successfully carried out above 
optical frequencies.




\subsection{Which telescope is best for what?}

Let us now put together the results from the previous subsections to investigate which telescope design
is most cost effective for various science goals. 

We will use the noise power spectrum $\Clnoise$ to quantify sensitivity.
We will begin our discussion 
focusing on only two parameters, the large-scale sensitivity $\Cnoise$ and the angular resolution $\theta$,
since the parametrization $\Clnoise=\Cnoise e^{\theta^2\l^2}$ is a reasonable approximation for many of the
telescope designs that we have discussed. We then turn to more general noise power spectra when 
discussing elongated FFTs, general interferometers and the issue of point source subtraction.

\subsubsection{Complementarity}

If we need a telescope with angular resolution $\theta$ and large-scale sensitivity $\Cnoise$,
then which design will meet out requirements at the lowest cost? 
The answer is summarized in \fig{ScopeComparisonFig} for a $\nu=150\>$MHz example.
First of all, we see that  SDTs, FFTTs and SITs and are highly complementary:
the cheapest solution is offered by SDTs for low resolution, 
FFTTs for high sensitivity $(\Cnoise)^{1/2}\simlt\theta\times 2 \mu$K, 
and standard interferometers or elongated FFTTs for high resolution 
$\theta\simlt (\Cnoise)^{1/2}/2 \mu$K, .

\subsubsection{Calculational details}

A few comments are in order about how these results were obtained.

For a single SRT, MFPT or FFTT, both the resolution and the sensitivity are determined by their area alone,
so as the area is scaled up, they each trace a line through the 
$(\theta,(\Cnoise)^{1/2})$ parameter space of \fig{ScopeComparisonFig}.
The cheapest way to attain a better sensitivity at the same resolution is simply to build multiple
telescopes of the same area (except for the FFTT, where cost$\simpropto A$, so that one might as well
build a single larger telescope instead and get extra resolution for free).
Since $\Cnoise\propto 1/N\Omega$, where $N$ is the number of telescopes whose images are averaged together,
the sensitivity of an FFTT with a given resolution can be matched by building 
$N=\Omega_{\rm FFTT}/\Omega$ telescopes,
where $N\sim A^{1/3}/\lambda^{2/3}$ for the MFPT and $N\sim A/\lambda^2$ for the SRT.
The cost relative to an FFTT of the same resolution and sensitivity  
thus grows as $A^{0.65}$ for MFPTs and as $A^{1.33}$ for SRT's.
The area below which single dish telescopes are cheaper depends strongly on wavelength; 
for the illustrative purposes of \fig{ScopeComparisonFig}, we have taken this to 
be $(10m)^2$ at 150 GHz based on crude hardware cost estimates for the GMRT \cite{GMRT} and MWA \cite{MWA} telescopes.
 
For regions to the right of the FFTT line in \fig{ScopeComparisonFig}, one has the option of either 
building a square (or circular) FFTT with unnecessarily high sensitivity to attain the required resolution, 
or to build an elongated FFTT or a conventional interferometer --- we return to this below, and argue that 
the latter is generally cheaper.

\subsubsection{How the results depend on frequency and survey details}

Although the \fig{ScopeComparisonFig} is for a specific example, these qualitative results hold more
generally. Survey duration, bandwidth, system temperature and sky coverage all merely rescale the 
numbers on the vertical axis, leaving the figure otherwise unchanged.
As one alters the observing wavelength, the resolution and sensitivity remains the same if one alters the other 
scales accordingly: $A\propto\lambda^2$, $c\tau\propto\lambda$, except that $\Tsys$ grows rapidly towards very low frequencies 
as the brightness temperature of synchrotron radiation exceeds the instrument temperature.
The cost depends strongly and non-linearly on frequency.
As discussed in \Sec{CostSec}, both the FFTT and digital SITs are currently feasible only below about 1 GHz, and 
and analog interferometry has not yet been successfully carried out above 
optical frequencies(?).

\subsubsection{The advantage of an FFTT over a single dish telescope}

The results above show that the FFTT can be thought of as simply a cheap single-dish telescope with a $180^\circ$ field of view.
Compared to single-dish telescope, the FFTT has two important advantages:
\begin{enumerate}
\item It is cheaper in the limit of a large collecting area, with the cost scaling roughly like $A$ rather than $A^{1.35}$ or more.
\item It has better power spectrum sensitivity even for fixed area $A$, because of a field of view that is larger by a factor 
between $(D/\lambda)^{2/3}$ and $(D/\lambda)^2$.
\end{enumerate}
An important disadvantage of the FFTT is that it currently only works below a about 1 GHz.
Even if it were not for this limitation, since the computational cost of interferometry depends on the number of resolution elements 
$N\sim\Omega/\theta^2$, which grows fast toward higher frequencies (as $\nu^{4/3}$ for the
MFPT and as $\nu^2$ for the FFTT), single-dish telescopes become comparatively more advantageous at higher frequencies.
However, as Moore's law marches on, the critical frequency where an FFTT loses out to an SDT should grow exponentially over time.

\subsubsection{The advantage of an FFT Telescope over a traditional correlating interferometer}

The results above also show that the FFTT can be thought of as a cheap maximally compact interferometer with a full-sky
primary beam. To convert a state-of-the-art interferometers such as MWA \cite{MWA}, LOFAR \cite{LOFAR}, PAPER\cite{PAPER}, 21CMA \cite{21CMA}
into an FFTT, one would need to do three things:
\begin{enumerate}
\item Move all antenna tiles together so that they nearly touch.
\item Get rid of any beamformer that ``points'' tiles towards a specific sky direction by adding relative phases 
to its component antennas, and treat each antenna as independent instead, thus allowing the array to image all sky directions 
simultaneously.
\item Move the antennas onto a rectangular grid to cut the correlator cost from $N^2$ to $N\log_2 N$.
\end{enumerate}
This highlights both advantages and disadvantages of the FFTT compared to traditional interferometers.
There are three important advantages:
\begin{enumerate}
\item It is cheaper in the limit of a large collecting area, with the cost scaling roughly like $A$ rather than $A^2$.
\item It has better power spectrum sensitivity even for fixed area $A$, because of a field of view that is larger than for an interferometer
whose primary beam is not full sky (because its array elements are either single-dish radio telescopes or antenna tiles that are 
pointed with beamformers).
\item The synthesized beam is as clean and compact as for a SDT, corresponding to something like a simple Airy pattern.
This has advantages for multifrequency point source subtraction as discussed below, and also for high fidelity mapmaking.
\end{enumerate}
The most obvious drawback of a square or circular FFTT 
is that the angular resolution is much poorer than what a traditional interferometer can deliver.
This makes it unsuitable for many traditional radio astronomy applications.
We discuss below how this drawback can be partly mitigated by a rectangular rather than square design.

A second drawback is the lack of flexibility in antenna positioning. Whereas traditional interferometry allows one to place the
antennae wherever it is convenient given the existing terrain, the construction of a large FFTT requires bulldozing.

\subsubsection{The advantage of a 2D FFTT over a 1D FFTT exploiting Earth Rotation}

There are two fundamentally different approaches to fully sampling a disk around the origin of the Fourier plane 
(usually referred to as the UV plane in the radio astronomy terminology): 
build a two-dimensional array (like a square FFTT) whose baselines cover this disk,
or build a more sparse array that fills the disk gradually, after adding together observations made at multiple times,
when Earth rotation has rotated the available baselines.
\Eq{CnoiseEq2} shows that, given a fixed number of antennas and hence a fixed collecting area, the former option gives 
lower $\Cnoise$ and hence more accurate power spectrum measurements as long as the angular resolution is sufficient. 
The reason is that the factor $\fcover$ in the denominator equals unity for the former case, and is otherwise smaller.
For a rectangular FFTT of dimensions $\Dmin\times\Dmax$, $B_\l^2\fcover$ depends on the angular scale
$\l$ and it is easy to show that 
\beq{fcoverEq}
B_\l^2\fcover \sim\left\{
\begin{tabular}{ll}
$1$				&for $\l\simlt\frac{\Dmin}{\lambda}$\\
$\frac{\Dmin}{\lambda\l}$	&for $\frac{\Dmin}{\lambda}\simlt\l\simlt\frac{\Dmax}{\lambda}$\\
$0$				&for $\l\simgt\frac{\Dmin}{\lambda}$
\end{tabular}
\right.
\eeq
In essence, making the telescope more oblong simply dilutes the same total amount of information out over a broader range of $\l$-space, thus giving 
poorer sensitivity on the angular scales originally probed.

What telescope configuration is desirable depends on the science goal at hand.
It has been argued \cite{Lidz07,21cmpars} that for doing cosmology with 21 cm tomography in the near term, it is best to make the telescope as 
compact as possible, \ie, to build a square or circular telescope. The basic origin of this conclusion is the result ``a rolling stone gathers no moss''
mentioned in \Sec{DClSec}: for power spectrum measurement, it is optimal to focus the efforts to make the signal-to-noise of order unity. 
The first generation of experiments have much lower signal-to-noise than this, and thus benefit from 
focusing on large angular scales and measuring them as accurately as possible rather than measuring a larger range of 
angular scales with even poorer sensitivity. Of course, 
none of these 1st generation telescopes were funded for 21cm cosmology alone, and their ability
to perform other science hinges on having better angular resolution, explaining why they were designed with less compact configurations.
Better angular resolution can also aid point source removal.

For other applications where high angular resolution required, an oblong telescope is preferable. 
An interesting proposal of this type is the higher-frequency mapping proposed Pittsburgh Cylinder telescope \cite{Peterson06,Chang07}, 
which is one-dimensional.
Instead rather omnidirectional antennas, it takes advantage of its one-dimensional nature by having a long cylindrical mirror, 
which increases the collecting area at higher frequencies.
This is advantageous because its goal is to map 21 cm emission at the lower redshifts (higher frequencies $\simgt$ 200 MHz) 
corresponding to the period after cosmic reionization, to detect neutral hydrogen in galaxies and use this to measure the 
baryon acoustic oscillation scale as a function of redshift.
If one wishes to perform rotation synthesis with an oblong or 1D FFTT, 
it will probably be advantageous to build multiple telescopes rotated relative to one another (say in an L-shaped layout, or like
spokes on a wheel), to reduce the amount of integration time needed to fill the UV plane.
Cross-correlating the $N$ antennas between the telescopes would incur a prohibitive $N^2$ computational cost,
so such a design with $T$ separate telescopes would probably need to discard 
all but of a fraction $1/T$ if the total information, corresponding to the intra-telescope baselines.

Another array layout giving higher resolution is to build an array whose elements consist of FFTTs placed far apart.
After performing a spatial FFT of their individual outputs, these can then be multiplied and inverse-transformed pairwise, and the resulting 
block coverage of the UV plane can be filled in by Earth rotation. As long as the number of separate FFTTs is modest, the extra numerical cost for this may 
be acceptable.

Above we discussed the tradeoff between different shapes for fixed collecting area.
If one instead replaces a $D\times D$ two-dimensional FFTT by a one-dimensional FFTT of length $D$ using rotation synthesis, 
then \eq{CnoiseEq2} shows that one loses sensitivity in two separate ways:
at the angular scale $\l\sim D/\lambda$ where the power spectrum error bar $\Delta C_\l$ from \eq{dClEq} is the smallest, 
one loses one factor of $D/\lambda$ from the drop in $\fcover$, and a second factor of $D/\lambda$
from the drop in collecting area $A$.
Another way of seeing this is to note that the available information scales as the number of baselines,
which scales as the square of the number of antennas and hence as $A^2$.
This quadratic scaling can also be seen in \eq{PSsensitivityEq2}: the total amount of 
information $(\Delta\phi)^{-2}$ scales as $A^2\Omega\tau\Delta\nu$, so whereas 
field of view, observing time and bandwidth help only linearly, area helps quadratically. This is because
we can correlate electromagnetic radiation at different points in the telescope, but not at different times, at different frequencies
or from different points in the sky.
The common statement that the information gathered scales as the etendu $A\Omega$ is thus true 
only at fixed $\l$; when all angular scales are counted, the scaling becomes $A^2\Omega$.

If in the quest of more sensitivity, one keeps lengthening an oblong or one-dimensional
FFT to increase the collecting area, one eventually hits a limit: 
the curvature of Earth's surface makes a flat $D\gg 10 km$ exceedingly costly, requiring instead telescope curving 
along Earth's surface and the alternative analysis framework mentioned above in \Sec{CostSec}.
If one desires maximally straightforward data analysis, one thus wants to 
grow the telescope in the other dimension to make it less oblong, as 
discussed in \Sec{CostSec}. This means that if one needs $\gg 10^4$ antennas for adequate 21 cm cosmology
sensitivity, one is forced to build a 2D rather than 1D telescope. For comparison, even the currently funded 
MWA experiment with its $512\times 4^2=8192$ antennas is close to this number.


One final science application where 2D is required is the study of transient phenomena
that vary on a time scale much shorter than a day, invalidating the static sky approximation
that underlies rotation synthesis. This was the key motivation behind the aforementioned
Waseda telescope \cite{Nakajima92,Nakajima93,Tanaka00}.

\section{Application to 21 cm tomography}
\label{21cmSec}

In the previous section we discussed the pros and cons of the FFTT telescope, 
and found that it's main strength is for mapping below 
about 1 GHz when extreme sensitivity is required.
This suggests that the emerging field of 21 cm tomography is an ideal first science application of 
the FFTT: it requires sky mapping in the sub-GHz frequency range, and the sensitivity requirements, especially to 
improve cosmic microwave background constraints on cosmological parameters, 
are far beyond what has been achieved in the past \cite{McQuinn06,Bowman07,21cmpars,LoebWyithe08}.

\subsection{21cm tomography science}

It is becoming increasingly clear that 21 cm tomography has great scientific potential
for both astrophysics \cite{Barkana01,ZaldaFurlanettoHernquist03,FurlanettoReview,Loeb06,Lidz07} 
and fundamental physics  \cite{BarkanaLoeb05,McQuinn06,Bowman07,21cmpars,LoebWyithe08}.
The basic idea is to produce a three-dimensional map of the matter distribution throughout our Universe 
through precision measurements of the redshifted 21 cm hydrogen line.
For astrophysics, much of the excitement centers around probing the cosmic dark ages and the 
subsequent epoch of reionization caused by the first stars.
Here we will focus mainly on fundamental physics, as this arguably involves both the 
most extreme sensitivity requirements and the greatest potential for funding extremely sensitive measurements.

\begin{figure}[ht]
\centerline{\epsfxsize=\figsize\epsffile{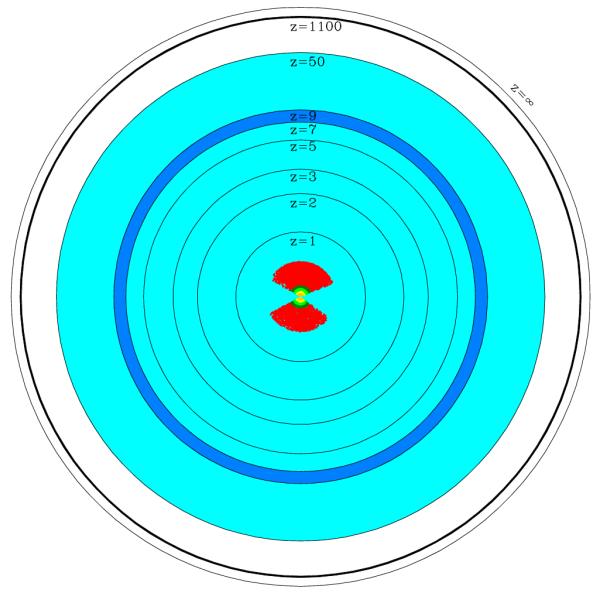}}
\caption{
21 cm tomography can potentially map most of our observable universe (light blue/gray), whereas the CMB probes mainly a thin 
shell at $z\approx 1100$ and current large-scale structure maps (here exemplified by the Sloan Digital Sky Survey and its luminous red galaxies)
map only small volumes near the center. Half of the comoving volume lies at $z>29$ (Appendix~\ref{GeometryAppendix}).
This paper focuses on the convenient $7\simlt z \simlt 9$ region (dark blue/grey).
\label{HubbleVolumeFig}
}
\end{figure}

\subsubsection{Three physics frontiers}

Future measurements of the redshifted 21 cm hydrogen line have the potential to probe hitherto unexplored regions of 
parameter space, pushing three separate frontiers: time, scale, and sensitivity.
Figure \ref{HubbleVolumeFig} shows a scaled
sketch of our observable Universe, our Hubble patch. It serves to show the
regions that can be mapped with various cosmological probes, and illustrates that 
the vast majority of our observable universe is still not mapped.
We are located at the center of the diagram. Galaxies (from the Sloan Digital Sky
Survey (SDSS) in the plot) map the distribution of matter in a three
dimensional region at low redshifts. Other popular probes like gravitational lensing, supernovae Ia,
galaxy clusters and the Lyman $\alpha$ forest are currently also limited to the small volume fraction corresponding to
redshifts $\simlt 3$ or less, and in many cases much less.
The CMB can be used to infer the
distribution of matter in a thin shell at the so-called ``surface of last
scattering", whose thickness corresponds to the width of the black circle at $z\sim 1100$ and thus covers only a tiny 
fraction of the total volume.
The region available for observation with the 21 cm line of
hydrogen is shown in light blue/grey. Clearly the 21 cm line of hydrogen has
the potential of allowing us to map the largest fraction of our observable universe
and thus obtain the largest amount of cosmological information.

At the high redshift end $(z\simgt 30)$ the 21 cm signal is relatively simple to
model as perturbations are still linear and ``gastrophysics'' related to stars and quasars is expected to be unimportant.
At intermediate times, during the
epoch of reionization (EOR) around redshift $z\sim 8$, the signal is
strongly affected by the first generation of sources of
radiation that heat the gas and ionize hydrogen. Modeling this era
requires understanding a wide range of astrophysical processes. At low
redshifts, after the epoch of reionization, the 21 cm line can be used to
trace neutral gas in galaxies and map the large scale distribution of
those galaxies.

\subsubsection{The time frontier}

\Fig{HubbleVolumeFig} illustrates that observations of the 21 cm line from the EOR and higher redshifts would map
the distribution of hydrogen at times where we currently have no other
observational probe, pushing the redshift frontier. Measurements of the
21 cm signal as a function of redshift will constrain the expansion
history of the universe, the growth rate of perturbations and the thermal
history of the gas during an epoch that has yet to be probed. 
\begin{itemize}
\item Tests of the standard model predictions for our cosmic  thermal history $T(z)$, expansion history $H(z)$ (which can be measured independently using 
both expansion and the angular diameter distances), and linear clustering growth.
\item Constraints on modified gravity from the above-mentioned measurements of $H(z)$ and clustering growth.
\item Constraints on decay or annihilation of dark matter particles, or any other long-lived relic, 
from the above-mentioned measurement of our thermal history \cite{Furlanetto06,Valdes07,Myers07}.
Here 21cm is so sensitive that even the expected annihilation of ``vanilla'' neutralino WIMP cold dark matter
may be detectable \cite{Myers07}. 
\item Constraints on evaporating primordial black holes from the thermal history measurement \cite{MackOstrikerRicotti08}.
\item Constraints on time-variation of fundamental physical constants such as the fine structure constant \cite{Khatri07}.
\end{itemize}

\subsubsection{The scale frontier}

\begin{figure}[ht]
\centerline{\epsfxsize=\figsize\epsffile{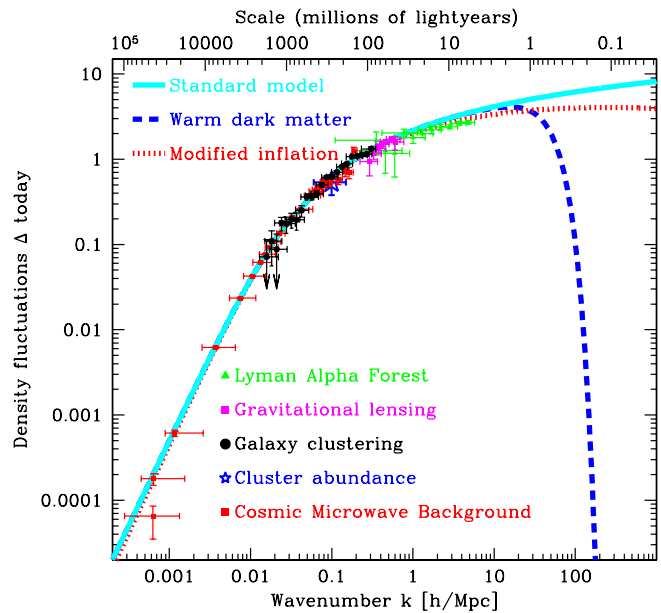}}
\caption{21 cm tomography can push the scale frontier far beyond that of current measurements
of cosmic clustering, potentially all the way down to the Jeans scale at the right edge
of the figure. This allows distinguishing between a host of alternative inflation and dark matter 
models that are consistent with all current data, for example a warm dark matter with mass
$14$ keV (dashed curve) or greater and inflation with a running spectral index 
more extreme than $dn_s/d\ln k=-0.03$ (dotted).
}
\label{ScaleFig}
\end{figure}

These observations can potentially push the ``scale frontier",
significantly extending the range of scales that are accessible to do
cosmology. This is illustrated in figure \ref{ScaleFig}, where the scales
probed by different techniques are compared to what is available in 21 cm.
Neutral hydrogen is a good probe of the small scales for two
separate but related reasons. First, one can potentially make
observations at higher redshifts, where more of the scales of interest are
in the linear regime and thus can be better modeled. 
Second, at early times in the history of our Universe,
hydrogen is still very cold and thus its distribution is expected to trace
that of the dark matter up to very small scales, the so-called Jeans
scale, where pressure forces in the gas can compete with gravity \cite{NaozBarkana05}. 

\begin{itemize}
\item Precision tests of inflation, since smaller scales provide a longer lever arm for constraining 
the spectral index and its running (illustrated in \fig{ScaleFig}) for the power spectrum of inflationary seed fluctuations \cite{21cmpars}
\item Precision tests of inflation by constraining small-scale non-Gaussianity \cite{Cooray08}.
\item Precision constraints on non-cold dark matter from probing galactic scales while they were still linear.
\end{itemize}

\subsubsection{The sensitivity frontier}

\begin{figure}[ht]
\centerline{\epsfxsize=\figsize\epsffile{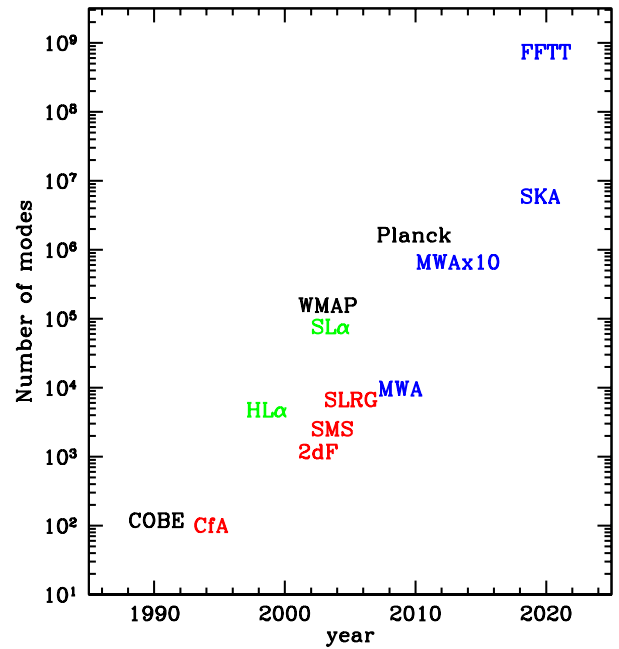}}
\caption{Number of modes measured with different cosmological probes. We show illustrative examples of galaxy redshift surveys (CfA, PsCz, 2dF, SDSS
main sample (SMS), SDSS Luminous red galaxies (SLRG)), CMB experiments (COBE, WMAP and Planck), Lyman-$\alpha$ forest measurements (using high
resolution spectra (HL$\alpha$) and SDSS spectra (SL$\alpha$)) and 21 cm experiments (MWA, an extension of MWA with ten times the collecting area,
the Square Kilometer Array (SKA) and a 1 km$^2$ FFTT). The number of modes is calculated from the constraints these experiments 
can place on the overall amplitude of the
power spectrum ($\delta P/P$) and then using the formula for Gaussian random fields $\delta P/P = \sqrt{2/N_{\rm modes}}$. }
\label{ModesFig}
\end{figure}

This combination of a large available volume with the presence of fluctuations on small scales that can be
used to constrain cosmology implies that the amount of information that at least in principle can be obtained 
with the 21 cm is extremely large. This can be illustrated by calculating the number of Fourier modes
available to do cosmology that can be measured with this technique. This number can be compared with the
number of modes measured to date with various other techniques such as galaxy surveys, the CMB, etc. In
figure \ref{ModesFig}, we show the number of modes measured by past surveys and some planned probes
including 21 cm experiments\footnote{Although the number of modes gives an
estimate of the statistical power of a survey, constraints on specific parameters will depend on how
strongly each of the power spectra varies as a function of the parameter of interest. Furthermore, when considering
probes such as the Lyman-$\alpha$ forest that probes modes in the non-linear regime, our numbers based on
the Gaussian formula overestimates the constraining power. In constructing this figure, only modes in the
linear regime $k<0.1\ {\rm h \ Mpc}^{-1}$ were included for galaxy surveys. These are the range of modes
that are typically used for doing cosmology. If the galaxy formation process becomes sufficiently well
understood it may become feasible to increase the number of useful modes. 
}. 
The figure illustrates a trend akin to Moore's law: exponential progress as a function of year. It is striking that
the improvement of the 1 km$^2$ FFTT over WMAP is comparable to that of WMAP over COBE. 
Moreover, the ultimate
number of modes available to be observed with 21 cm tomography is dramatically larger still, upward of $10^{16}$,
so although many practical issues will
most certainly limit what can be achieved in the near future, the ultimate potential is vast.

The FFTT sensitivity improvement translates into better measurement accuracy for many of the usual cosmological parameters.
It has been shown that
even the limited redshift range $7\simlt z\simlt 9$ (dark shading in \fig{HubbleVolumeFig}) has the potential to
greatly improve on cosmic microwave background constraints from WMAP and Planck:
it could improve the sensitivity to spatial curvature and neutrino masses by up
to two orders of magnitude, to $\Delta\Omega_k\approx 0.0002$ and
$\Delta m_\nu\approx 0.007$ eV, and give a $4\sigma$ detection of the
spectral index running predicted by the simplest inflation models \cite{21cmpars}.
Indeed, it may even be possible to measure three individual neutrino masses
from the scale and time dependence of clustering \cite{21cmpars,PritchardPierpaoli08}.


Measuring the 21 cm power spectrum and using it to constrain physics and astrophysics does not require pushing the noise level down to 
the signal level, since the noise can be averaged down by combining many Fourier modes probing the same range of scales.
This is analogous to how the COBE satellite produced the first measurement of the CMB power spectrum even though 
individual pixels in its sky maps were dominated by noise rather than signal \cite{Smoot92}.
Further boosting the sensitivity to allow imaging (with signal-to-noise per pixel exceeding unity)
allows a number of improvements:
\begin{itemize}
\item Improving quantification, modeling and understanding of foregrounds and systematic errors
\item Pushing down residual foregrounds with better cleaning (like in the CMB field, the residual foreground level after cleaning is likely to be 
comparable to the noise level)
\item Enabling power spectrum and non-Gaussianity estimation after masking out ionized bubbles, 
thus greatly reducing the hard-to-model ``gastrophysics'' contribution
\item Constraining small-scale properties of dark matter by using 21 cm maps as backgrounds 
for gravitational lensing experiments that could detect the presence of 
dark substructure in lower redshifts halos \cite{Zahn06,Metcalf06,Hilbert07} 
\item Pushing to higher redshift where the physics is simpler  
\end{itemize}

\subsection{The cost of sensitivity}

There is thus little doubt that sensitivity improvements can be put to good use.
\Eq{PnoiseEq} implies that the high-redshift frontier in particular 
has an almost insatiable appetite for sensitivity:
since $\lambda\propto (1+z)$, $y\propto (1+z)^{1/2}$, $d_A$ depends only weakly on $z$, and 
the diffuse synchrotron foreground that dominates $\Tsys$ at 
low frequencies scales roughly as $\nu^{-2.6}\propto (1+z)^{2.6}$ in the cleanest parts of 
the sky for $50\simlt\nu\simlt 200$ MHz \cite{gsm}, \eq{PnoiseEq} gives a sensitivity
\beq{HighzEq}   
\delta T\simpropto\left[k^3\Pnoise\right]^{1/2}
\propto\frac{k^{3/2}(1+z)^{3.85}\fsky^{1/2}}{(A\Omega\fcover\tau)^{1/2}}.
\eeq
if the observing time and field of view is held fixed (like for the FFTT).
Pushing from $z=9$ to $z=20$ with the same sensitivity thus requires increasing the collecting area by a factor around 300.
This would keep the signal-to-noise level roughly the same if the 
21 cm fluctuation amplitude is comparable and peaks at similar angular scales at the two redshifts, as suggested by the calculations 
of \cite{Pritchard08}. \Eq{HighzEq} shows that imaging smaller scales is expensive too, with an order of magnitude
smaller scales (multiplying $k$ by 10) requiring a thousandfold increase in collecting area.

\clearpage 
\begin{figure}[ht]
\centerline{\epsfxsize=\figsize\epsffile{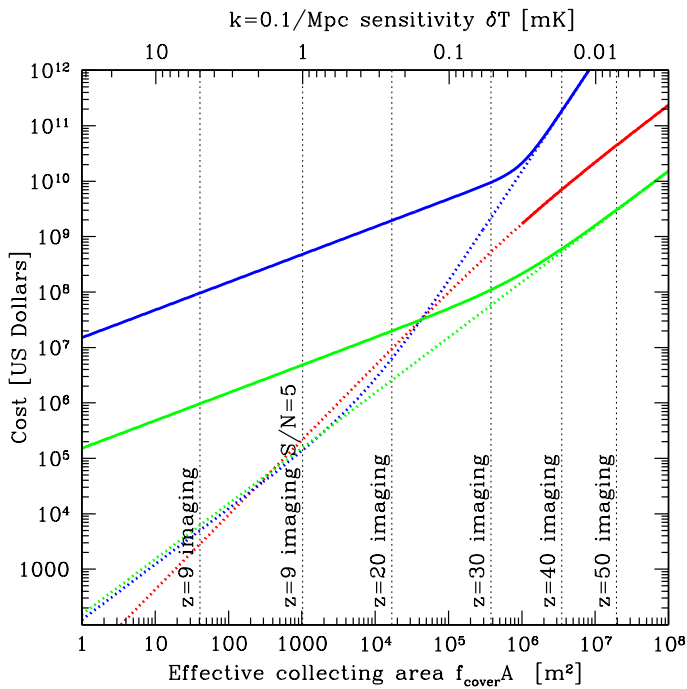}}
\caption{The rough hardware cost in 2008 US Dollars of attaining various sensitivities at the 
$k=0.1/{\rm Mpc}$ scale with the FFTT telescope (green curves),
a maximally compact regular interferometer (blue curves) and a single-dish telescope (red curve)
always pointing towards the same patch of sky ($4\pi\fsky\Omega$).
The dashed curves have angular resolution poorer than $\l=500$ at redshift $z=9$; 
for the SIT and FFTT, this resolution can be achieved by making the telescope array
oblong a higher cost (solid curves), since the area must be increased to compensate for the drop in $\fcover$. Note that cost is a function of $A$ only so to plot is as a function of sensitivity $A\fcover$ a design dependent relation between $A$ and $\fcover$ is required. 
}
\label{CostFig1}
\end{figure}

\begin{figure}[ht]
\centerline{\epsfxsize=\figsize\epsffile{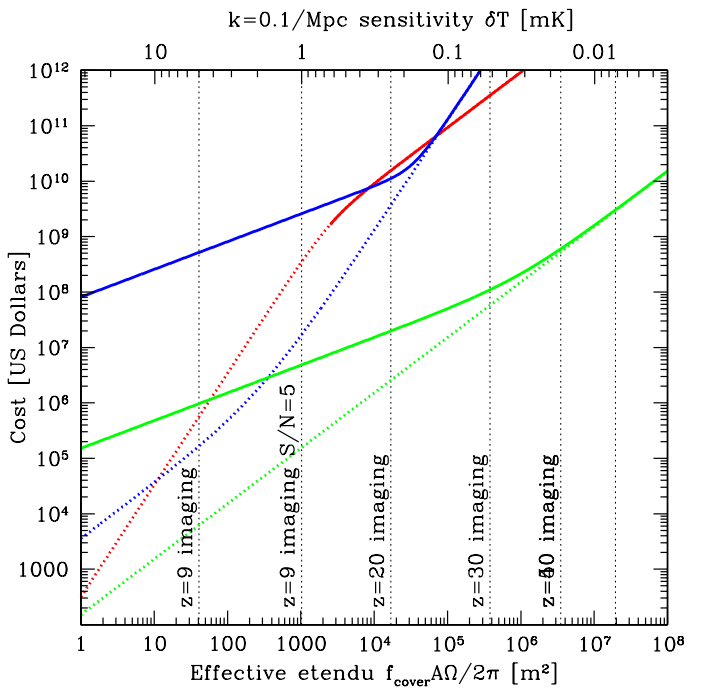}}
\caption{Same as previous figure, but when half the sky is 
mapped ($\fsky=2\pi$). The standard interferometric telescope (SIT) and single dish 
maximal focal plane telescope (MFPT) take an additional cost his here, needing to 
further increase the area to compensate for the drop in field of view $\Omega$ with $A$.
}
\label{CostFig2}
\end{figure}

Figures~\ref{CostFig1} and~\ref{CostFig2} illustrate the rough cost of attaining the sensitivity 
levels required for various physics milestones mentioned above.
Our cost estimates are very crude, and making more accurate ones 
would go beyond the scope of the present paper, but the qualitative scalings
seen in the figures should nonetheless give a good indication of how 
the different telescope designs complement each 
other.\footnote{For interferometer arrays, we use the following hardware cost estimate loosely 
based on the MWA hardware budget \cite{MWA}:
\$1M$\times(A/8000{\rm m}^2)\beta+$\$1M$\times(A/8000{\rm m}^2)^\gamma$,
where $(\beta,\gamma)=(1.2,1)$ for the FFTT and  $(\beta,\gamma)=(1,2)$ for a conventional interferometer.
The first term corresponds to per-antenna costs (with $\beta$ reflecting
the extra construction cost related to land leveling {\etc}), and the second term corresponds to 
the computational cost.
For a single dish, we assume a hardware cost 
\$0.4M$\times[A/(1600{\rm m}^2)]^{1.35}$
based on Wilkinson's scaling \cite{Wilkinson90} from the GMRT budget \cite{GMRT}.
}
For our estimates, we have assumed $\tau=4000$ hours of observation
with a system temperature $\Tsys=200$K$\times[(1+z)/10]^{2.6}$.
We assume the cosmic signal to be of order $\delta T=5$mK at the redshifts of interest
\cite{Pritchard08}.
Our baseline estimates are an observing frequency of $142$ MHz, corresponding to 21 cm emission at redshift $z=9$. 

Figure~\ref{CostFig1} is for the case when all we care about is sensitivity, not how large a sky area is mapped with
this sensitivity. We thus keep the telescope pointing at the same sky patch and get
$4\pi\fsky\Omega$, so \eq{HighzEq} gives a sensitivity
$\delta T\propto(k^{3/2}(1+z)^{3.85})/(A\fcover)^{1/2}$.
For a fixed spatial scale $k$ and redshift $z$, the
sensitivity thus depends only on the collecting 
effective area $\fcover A$ plotted on the horizontal axis.
The solid curves in the figure all have maximally compact configurations with 
$\fcover=1$, corresponding to angular resolution $\l\sim A^{1/2}/\lambda$.
The lines are dotted where this resolution $\l<500$ for the baseline wavelength
$\lambda=2.1$m. If we insist on the higher resolution $\l=500$, we can achieve this goal
by making the FFT or SIT oblong or otherwise sparse, with 
$\fcover\sim A/(\lambda\l)^2\propto A$, so in this regime,
$(A\fcover)\propto A^2$ and hence $A\propto (A\fcover)^{1/2}$, and 
this area in turn determines the cost --- this is why the solid curves in \fig{CostFig1}
lie above the corresponding dotted ones.

Figure~\ref{CostFig2} is for the case when we want a map of a fixed area
(WMAP-style), in this case covering half the sky ($\fsky=0.5$), 
so \eq{HighzEq} gives a sensitivity
$\delta T\propto(k^{3/2}(1+z)^{3.85})/(A\Omega\fcover)^{1/2}$.
For a fixed spatial scale and redshift, the
sensitivity thus depends only on the collecting 
effective etendu $\fcover A\Omega$ plotted on the horizontal axis.
Since $\Omega$ drops with area for both the SRT and MFPT, 
in order to boost sensitivity, these telescopes now need an extra area boost to make up for the drop in $\Omega$.
Although an MFPT has cost$\propto A^{1.35}$, it also has $\Omega\propto A^{-1/3}$, so that $A\Omega\propto A^{2/3}$ and
the cost $\propto (A\Omega)^{1.35\times 3/2}\approx (A\Omega)^2$.
Once $A$ is large enough to give sufficient resolution ($A\simgt\lambda^2\l^2$) 
it becomes smarter to simply build multiple telescopes, giving cost $\propto A$. 

For comparison, we have indicated some sensitivity benchmarks as vertical lines.
\Eq{HighzEq} shows that $\delta T\simpropto k^{3/2}(1+z)^{3.85}$; this redshift scaling is illustrated by these
vertical lines. Additional sensitivity can also be put to good use for probing smaller scales, since 
an order of magnitude change in $k$ corresponds to three orders of magnitude on the horizontal axis.


\subsection{21 cm foregrounds}

Aside from its extreme sensitivity requirements, another unique feature of 21 cm cosmology is the magnitude of its foreground problem:
it involves mapping a faint and diffuse cosmic signal that needs to be separated from
foreground contamination that is many orders of magnitude brighter \cite{FurlanettoReview,Santos07,gsm}, requiring extreme sensitivity and beam control.
Fortunately, the foreground emission (mainly synchrotron radiation) has a rather smooth frequency spectrum, while the cosmological signal 
varies rapidly with frequency (corresponding to variations in physical conditions along the line of sight).
Early work on 21cm foregrounds \cite{21cm,Morales06,Jelic08} has indicated that this can be exploited to clean out the foregrounds
down to an acceptable level, effectively by high-pass filtering the data cube in the frequency direction. 

However, these papers have generally not treated the additional complication that the synthesized beam $\Wcaret(\theta_x,\theta_y)$
is frequency dependent, dilating like $\lambda$, which means that when raw sky maps at two different frequencies cannot be readily 
compared. For a single-dish telescope or an FFTT, the synthesized beam is compact and simple enough that this complication can be
modeled and remedied exactly (say by convolving maps at all frequencies to have the same resolution before foreground cleaning), but for 
a standard interferometer, complicated low-level ``frizz'' extending far from the central parts of the synthesized beam appears to make this unfeasible
at the present time.
Recent work \cite{BowmanThesis,Bowman08,21cmps} has indicated that this is a serious problem: whereas the foreground emission from our own galaxy 
is smooth enough that these off-beam contributions average down to low levels, emission from other galaxies appears as point sources to which the
telescope response varies rapidly with frequency because of the beam dilation effect.
The ability to mitigate this problem is still subject to significant uncertainty \cite{21cmps}, and 
may therefore limit the ultimate potential of 21 cm cosmology with a conventional interferometer.
The ability to deal with foreground contamination is thus another valuable advantage of the FFT Telescope.

\section{Conclusions}
\label{ConcSec}


We have presented a detailed analysis of an all-digital telescope design where mirrors are replaced by fast Fourier transforms,
showing how it complements conventional telescope designs. 
The main advantages over a single dish telescope are cost and orders of magnitude larger field-of-view, translating into dramatically 
better sensitivity for large-area surveys.
The key advantages over traditional interferometers are cost (the correlator computational cost for an
$N$-element array scales as $N\log_2 N$ rather than $N^2$) and a compact synthesized beam.
These traits make the FFT Telescope ideal for applications where the angular resolution requirements are modest while those on sensitivity are extreme.
We have argued that the emerging field of 21 cm tomography could provide an ideal first application of a very large FFT Telescope, since
it could provide massive sensitivity improvements per dollar as well as mitigate the off-beam point source foreground problem with its 
clean beam.

\subsection{Outstanding challenges}

There are a number of interesting challenges and design questions that would need to be addressed
before building a massive FFT Telescope for 21 cm cosmology. For example:
\begin{enumerate}
\item To what extent can the massive redundancy of an FFT Telescope (where the same baseline is typically measured by $\sim N_a^{1/2}$ 
independent antenna pairs) be exploited to calibrate the antennas against one another in a computationally feasible way?
\item To what extent, if any, are more distant antennas outside the FFTT needed to resolve bright point sources and calibrate the FFTT antennas?
\item After calibration, how do gain fluctuations in the individual array elements affect the noise properties of the recovered sky map?
\item How do variations in primary beam $\B(\khat)$ from \eq{ResponseEq1} from between individual antennas affect the properties of the recovered sky map?
\item How many layers of dummy antennas are needed around the active instrumented part of the array to ensure 
that the beam patterns of all utilized antennas are sufficiently identical?
\item What antenna design is optimal for a particular FFT Telescope science application, maximizing gain in the relevant 
frequency range? The limit of an infinite square 
grid of antennas on an infinite ground screen is quite different from the limit of a single isolated antenna,
and modeling mutual coupling effects becomes crucial when computing the primary beam $\B(\khat)$ from \eq{ResponseEq1}
\item What unforeseen challenges does the FFT Telescope entail, and how can they be overcome?
\item Can performing the first stages of the spatial FFT by analog means (say connecting adjacent $2\times 2$
or $4\times 4$ antenna blocks with Butler matrices \cite{Butler61}) lower the effective system temperature in parts 
of the sky with overall lower levels of synchrotron emission? 
\end{enumerate}
Answering these questions will require a combination of theoretical and experimental work.
The authors are currently designing a small FFTT prototype with a group of radio astronomy colleagues to address these questions
and to identify unforeseen obstacles.


\subsection{Outlook}

Looking further ahead, we would like to encourage theorists to think big and look into what additional physics may be learned from 
the sort of massive sensitivity gains that an FFTT could offer, as this can in turn increase the motivation for hard work on experimental 
challenges like those listed above.

Perhaps in a distant future, almost all telescopes will be FFT Telescopes, simultaneously observing light of all wavelengths from all directions.
In the more immediate future, as Moore's law enables FFTT's with higher bandwidth, cosmic microwave background polarization may be an interesting 
application besides 21 cm cosmology. By using an analog frequency mixer to extract of order a GHz of bandwidth in the CMB frequency range
(around say 30 GHz or 100 GHz), it would be possible to obtain a much greater instantaneous sky coverage than current CMB experiments provide, and this 
gain in $\Omega$ could outweight the disadvantage of lower bandwidth $\Delta\nu$ in \eq{CnoiseEq2} to provide overall better sensitivity.
The fact that extremely high spectral resolution would be available essentially for free may also help ground-based measurements, allowing exploitation of
the fact that some atmospheric lines are rather narrow.

\bigskip
{\bf Acknowledgements:}
The authors wishes to thank Michiel Brentjens, Don Backer, Angelica de Oliveira-Costa, Ron Ekers, 
Jacqueline Hewitt, Mike Jones, Avi Loeb, Adrian Liu, Ue-Li Pen, Jeff Peterson,
Miguel Morales, Daniel Mitchell, James Moran, Jonathan Rothberg, Irwin Shapiro, Richard Thompson and an anonymous referee for helpful comments, and Avi Loeb in particular for
encouragement to finish this manuscript after a year of procrastination.
This work was supported by NASA grants NAG5-11099 and NNG 05G40G,
NSF grants AST-0134999 and AST-05-06556, 
a grant from the John Templeton foundation
and fellowships from the David and Lucile
Packard Foundation and the Research Corporation.

\appendix

\section{Polarization issues}
\label{PolarizationSec}

The Stokes matrix $\S$ defined by \eq{SdefEq}
is related to the usual Stokes parameters $I,Q,U,V$ by
\beq{StokesEq}
\S=\frac{1}{2}\left(
\begin{tabular}{cc}
$I+Q$	&$U-iV$\\
$U+iV$	&$I-Q$
\end{tabular}
\right)
=\frac{1}{2}\vv\cdot\vsigma.
\eeq
In the dot product, 
\beq{StokesVectorEq}
\vv\equiv\left(
I\quad Q\quad U\quad V
\right)
\eeq
and 
\beq{PauliEq}
\vsigma\equiv\left\{
\left(
\begin{tabular}{cc}
$1$	&$0$\\
$0$	&$1$
\end{tabular}
\right),
\left(
\begin{tabular}{rr}
$1$	&$0$\\
$0$	&$-1$
\end{tabular}
\right),
\left(
\begin{tabular}{cc}
$0$	&$1$\\
$1$	&$0$
\end{tabular}
\right),
\left(
\begin{tabular}{rr}
$0$	&$-i$\\
$i$	&$0$
\end{tabular}
\right)
\right\},
\eeq
contains the four Pauli matrices.
As usual, $I$ denotes the total intensity, $Q$ and $U$ quantify the 
linear polarization
and $V$ the circular polarization (which normally vanishes for astrophysical sources).
It is easy to invert \eq{StokesEq} to solve for the Stokes parameters:
\beq{StokesSolutionEq}
\vv=\tr\{\sigma\cdot\S\}.
\eeq

An annoying but harmless nuisance when dealing with large-area polarization maps is the well-known fact that 
``you can't comb a sphere'', \ie, that there is no global choice of reference vector to define
the Jones vector and the Stokes parameters $(Q,U)$ all across the sky.
In practice, it never matters until at the very last analysis step, since one can collect the
data $\d_i$ and reconstruct both $\Shat_B$ and $\S_B$ without worrying about this issue.
To compute $\B$ and solve for the Stokes parameters, any convention for defining the Stokes parameters will suffice, 
even one involving separate schemes for a number of partially overlapping sky patches; it is easy to see that
the choice of convention has no effect on the accuracy or numerical stability of the inversion method.

\section{Cosmic geometry}
\label{GeometryAppendix}

In this Appendix, we derive equations\eqn{GeometryApproxEq}
and~\eqn{GeometryApproxEq2}.
For a flat universe (which is an excellent approximation for ours \cite{lrg,wmap5pars}),
the comoving angular diameter distance is given by \cite{KolbTurnerBook}
\beq{dAEq}
d_A(z)=\int_0^z \frac{c\,dz'}{H(z')},
\eeq
where 
\beq{EdefEq}
H(z)=H_0\sqrt{\Om(1+z)^3 + \Ol},
\eeq
where $\Ol=1-\Om$.
The second term in the square root becomes negligible for 
$z\gg (\Ol/1-\Om)^{1/3}-1\approx 0.4$ for $\Om=0.25$ \cite{wmap5pars}, which gives \eq{GeometryApproxEq2}.
The dark energy density is completely negligible at the high redshift regime 
relevant to 21 cm cosmology
also in most models where this density evolves with time. For such high redshifts, 
we can therefore approximate \eq{dAEq} as follows:
\beqa{dAEq2}
d_A(z)&=&\int_0^\infty \frac{c\,dz'}{H(z')} - \int_z^\infty \frac{cH_0^{-1}dz'}{\Om^{1/2}(1+z')^{3/2}}\nonumber\\
      &\approx&H_0^{-1}c\left[3.56 - \frac{4}{(1+z)^{1/2}}\right]
\eeqa
for $\Om=0.25$, which gives \eq{GeometryApproxEq}.
The accuracy of \Eq{dAEq2} better than 1\% for $z>2.2$, \ie, better that with which the relevant cosmological parameters
have currently been measured.

\Eq{dAEq2} shows that, surveying our observable universe as illustrated in \fig{HubbleVolumeFig}, 
we reach half the comoving distance at
$z\approx (4/(3.56\times 0.5))^2 - 1\approx 4$
and half the comoving volume at 
$z\approx\{4/[3.56\times (1 - 0.5^{1/3})]\}^2 - 1\approx 29$.
.

\section{Field-of-view estimates}
\label{FOVappendix}


In this appendix, we derive the restriction on the field of view for a single dish telescope. 
Consider a parabolic mirror of height $z$ given by:
\beq{zDefEq}
z={x^2 + y^2 \over R}
\eeq
where $x$ and $y$ are the coordinates in the plane of the ground and $R$ determines the radius of curvature.  The mirror has a diameter $D$ such that 
\beq{DdefEq}
x^2 + y^2 < {D^2 \over 4}.
\eeq
We consider radiation initially traveling with wave vector $\khat= k ( \sin \theta, 0, \cos \theta)$ with $k=2\pi/\lambda$. We will calculate the phase of the radiation that scatters at the location $(x,y,z)=(\rho \cos \phi, \rho \sin \phi, \rho^2 / R)$ on the surface of the mirror and 
then arrives at a detector located at $(x_f,y_f,z_f)$. For simplicity we will consider a point in the mirror with $y=0$ so that $\phi=0$ and then also set $y_f =0$. 
After some simple algebra one obtains the following expression for  the phase $\psi$: 
\beq{psi_eq}
\psi
=k\left[\sqrt{(x_f-\rho)^2 + \left(z_f - {\rho^2 \over R}\right)^2} - {\rho (\rho \cos \theta + R \sin \theta) \over {R}}\right]. 
\eeq  
Because of the parabolic shape chosen for the mirror, the phase of radiation coming with normal incidence ($\theta=0$) comes to a perfect focus at $x_f=0$, $z_f=R/4$. 
By perfect focus we mean that the phase 
$\psi(x_f,z_f,\rho)$ is independent of $\rho$ for $x_f=0$, $z_f=z/4$, $\theta=0$
For radiation incident at an angle, there will be no point in space where one can locate the detector so that the radiation reflected everywhere in the mirror will be in phase. We will find the field of view of the telescope by demanding that the phase difference between radiation incident in different parts of the telescope be less than a radian at the location of the detector.  

To obtain a formula, we expand $\psi$ in a Taylor series as a function of $\rho$. By choosing $x_f$ and $z_f$, 
we can make  the terms linear and quadratic in $\rho$ vanish, but the cubic term will in general be non-zero, except for normal incidence. For a given telescope diameter we will then find the field of view by demanding that the cubic contribution to the phase be smaller than a radian. The Taylor series of $\psi$ is: 
\beqa{psiExpansionEq}
k^{-1}\psi&\approx&
\sqrt{{x_f}^2+{z_f}^2}
- \left(\frac{{x_f}}{\sqrt{{x_f}^2+{z_f}^2}}+\sin \theta \right) \rho  \nonumber \\
&+& \frac{\left(\frac{{z_f} \left((R-2 {z_f}) {z_f}-2 {x_f}^2\right)}{\left({x_f}^2+{z_f}^2\right)^{3/2}}-2 \cos \theta \right) \rho ^2}{2 R} \nonumber \\
&+&  \frac{{x_f} {z_f} \left((R-2 {z_f})
   {z_f}-2 {x_f}^2\right) \rho ^3}{2 R
   \left({x_f}^2+{z_f}^2\right)^{5/2}} + \cdots
\eeqa
By choosing
\beq{xfEq}
x_f=-{z_f} \tan \theta\approx - {z_f} \theta, 
\eeq
we can eliminate the term linear in $\rho$, and by choosing
\beq{zfEq}
z_f=\frac{R}{8} (\cos 2 \theta  +1)\approx \frac{R}{4}
\eeq
the quadratic one. Thus we get
\beq{psiEq2}
\psi\approx k \  \left(\frac{1}{4} R \cos \theta -\frac{4  \rho ^3 \sin \theta.
   }{R^2} +\cdots \right).
\eeq
For small values of $\theta$, demanding that $\psi$ changes by less than a radian as we 
move from the center to the edge of the telescope, and using $k=2\pi/\lambda$,
 we obtain, 
\beq{th1Eq}
\theta < \frac{R^2 \lambda }{D^3 \pi } =  \frac{R^2}{D^2 \pi }  \times  \frac{\lambda }{D} .
\eeq 
Thus by increasing the radius of curvature $R$, one can increase the field of view. 
In fact, $(R/D)^2$ basically gives the number of resolution elements 
in each linear dimension in the focal plane. 

The upper bound on the size on the curvature radius comes from demanding that the focal plane 
not cover the entire telescope. Using equations (\ref{xfEq}) and (\ref{zfEq}) and demanding that the size of the focal plane be smaller than $D/2$ 
(a very conservative assumption), we get another constraint on the field of view: 
\beq{th2Eq}
\theta < \frac{2 D}{R}.
\eeq 
While the size of the field of view increases with $R$ in (\ref{th1Eq}), it decreases with $R$ 
in (\ref{th2Eq}) and thus the largest field of view is obtained when both constraints are equal, 
and corresponds to
\beq{Req2}
R=\left(\frac{2 \pi D}{\lambda}\right)^{1/3} D \  ; \quad \theta < \left(\frac{4 \lambda}{\pi D}\right)^{1/3}.
\eeq
The inequality that we have been derived can be pushed somewhat with clever multi-mirror designs 
(for example, the optical large synoptic telescope uses three mirrors \cite{LSST}).
In contrast, radio telescopes typically use only one mirror.
In this case, the value of $R$ required to attain the maximal field of view that we have derived is a factor $(2\pi D/\lambda)^{1/3}$ 
larger than $D$ and can thus get very large for sufficiently small wavelengths. 
Mechanical constraints can make building such a radio telescope impractical as the focal plane would be very far away from the telescope, 
making the upper bound $\theta\simlt (\lambda/D)^{1/3}$ that we have derived for the field of view a rather conservative one.





\vskip-1.0cm

\begin{thebibliography}{99}

\bibitem{MWA}
\rn\url{http://www.haystack.mit.edu/ast/arrays/mwa/}

\bibitem{LOFAR}
\rn\url{http://www.lofar.org/}

\bibitem{21CMA}
\rn\url{http://web.phys.cmu.edu/~past/}, formerly known as {\it PaST}.

\bibitem{PAPER}
\rn\nn Backer D, private communication (2008)

\bibitem{GMRT}
\rn\url{http://gmrt.ncra.tifr.res.in/}


\bibitem{GMRT2}
\rfprep\nn Pen U, \nn Chang T, \nnn Peterson J B, Roy J, \nn Gupta Y\multiand\nn Bandura K;2008;{arXiv:0804.2501 [astro-ph]}

\bibitem{SKA}
\texttt{http://www.skatelescope.org/}


\bibitem{Butler61}
\rf\nn Butler J\dualand Lowe R;1961;Electronic Design;9;170-173

\bibitem{May84}
\rf\nn May J, \nn Reyes F, \nn Aparici J, \nn Bitran M, \nn Alvarez H\multiand\nn Olmos F;1984;A\&A;140;377

\bibitem{Nakajima92}
\rf\nn Nakajima J {\etal};1993;PASJ;44L;35

\bibitem{Nakajima93}
\rf\nn Nakajima J {\etal};1993;PASJ;45;477

\bibitem{Tanaka00}
\rf\nn Tanaka N {\etal};2000;PASJ;52;447

\bibitem{Pen04}
\rf\nn Pen U;2004;New Astron.;9;417

\bibitem{Takeuchi05}
\rf\nn Takeuchi H {\etal};2005;PASJ;57;815
	     
\bibitem{Peterson06}
\rfprep\nnn Peterson J B, \nn Bandura K\multiand\nnn Pen U L;2006;astro-ph/0606104

\bibitem{Chang07}
\rf\nn Chang T, \nn Pen U, \nnn Peterson J B\multiand\nn McDonald P;2008;PRL;100;091303

\bibitem{SKAprivComm}
\rn After presenting this work at a May 2008 conference, we were told that the Square Kilometer Array collaboration is 
considering similar ideas for the central part of their proposed telescope (Jeffrey B.~Peterson, private communication)





\bibitem{Barkana01}
\rf\nn Barkana R\dualand\nn Loeb A;2001;Phys.Rept.;349;125

\bibitem{ZaldaFurlanettoHernquist03}
\rf\nn Zaldarriaga M, \nn Furlanetto S\multiand\nn Hernquist L;2004;ApJ;608;622

\bibitem{FurlanettoReview}
\rf\nn Furlanetto S, \nnn Oh S P\multiand\nn Briggs F;2006;Phys.Rept.;433;181

\bibitem{Loeb06}
\rf\nn Loeb A;2006;Sci.Am.;295;{46, astro-ph/0702298}

\bibitem{Santos07}
\rf\nnn Santos M G, \nn Amblard A, \nn Pritchard J, \nn Trac H, \nn Cen R\multiand\nn Cooray A;2008;ApJ;689;1-16

\bibitem{Pritchard08}
\rf\nnn Pritchard J R\dualand\nn Loeb A;2008;PRD;78;103511

\bibitem{21cmpars}
\rf\nn Mao Y, \nn Tegmark M, \nn McQuinn M, \nn Zahn O\multiand\nn Zaldarriaga M;2008;PRD;78;023529

\bibitem{ThompsonBook}
\rfbook\nnn Thompson A R\dualand\nnn Moran J M\multiand\nnn Swenson G W;2001;Interferometry and Synthesis in Radio Astronomy, 2nd Ed.;Wiley;{New York}

\bibitem{wmap5beams}
\rf\nnn Hill R S;2009;ApJS;180;246



\bibitem{Knox95}
\rf L. Knox;1995;Phys. Rev. D;52;4307

\bibitem{wiener}
\rf M. Tegmark and G. Efstathiou;1996;MNRAS;281;1297

\bibitem{lrg}
\rf\nn Tegmark M {\it etal};2006;PRD;74;123507

\bibitem{wmap5pars}
\rf\nn Dunkley J;2009;ApJS;180;306

\bibitem{Scott94}
\rf\nn Scott D, \nn Srednicki M\multiand\nn White M;1994;ApJ;421;L5

\bibitem{strategy}
\rf\nn Tegmark M;1997;Phys. Rev. D;56;4514

\bibitem{Wilkinson90}
\rn\nnn Wilkinson P N, in {\it Radio Interferometry: Theory, Techniques, and Applications: Proceedings of the 131st 
IAU Colloquium}, Astronomical Society of the Pacific (1990)

\bibitem{Legg98}
\rf\nnn Legg T H;1998;A\&AS;130;369

\bibitem{Lidz07}
\rf\nn Lidz A, \nn Zahn O, \nn McQuinn M, \nn Zaldarriaga M\multiand\nn Hernquist L;2008;ApJ;680;962




\bibitem{BarkanaLoeb05}
\rf\nn Barkana R\dualand\nn Loeb A;2005;ApJL;624;L65

\bibitem{Bowman07}
\rf\nnn Bowman J D, \nnn Morales M F\multiand\nnn Hewitt J N;2007;ApJ;661;1

\bibitem{McQuinn06}
\rf\nn McQuinn M, \nn Zahn O, \nn Zaldarriaga M, \nn Hernquist L\multiand\nnn Furlanetto S R;2006;ApJ;653;815

\bibitem{LoebWyithe08}
\rf\nn Loeb A\dualand\nn Wyithe S;2008;PRL;100;161301


\bibitem{Furlanetto06}
\rf\nn Furlanetto S, \nnn Oh P O\multiand\nn Pierpaoli E;2006;PRD;74;103502

\bibitem{Valdes07}
\rf\nn Valdes M, \nn Ferrara A, \nn Mapelli M\multiand Ripamonti E;2007;MNRAS;377;245

\bibitem{Myers07}
\rf\nn Myers Z\dualand\nn Musser A;2008;MNRAS;384;727

\bibitem{MackOstrikerRicotti08}
\rf\nnn Mack K J, \nnn Ostriker J P\multiand\nn Ricotti M;2008;ApJ;680;829

\bibitem{Khatri07}
\rf\nn Khatri R\dualand\nnn Wandelt B D;2007;PRL;98;111301

\bibitem{NaozBarkana05}
\rf\nnn Naoz S\dualand\nn Barkana R;2005;MNRAS;362;1047

\bibitem{Cooray08}
\rf\nn Cooray A, \nn Li C\multiand\nn Melchiorri A;2008;PRD;77;103506

\bibitem{PritchardPierpaoli08}
\rf\nnn Pritchard J R\dualand\nn Pierpaoli E;2008;PRD;78;065009

\bibitem{Smoot92}
\rf\nnn Smoot G F {\etal};1992;ApJL;396;L1

\bibitem{Zahn06}
\rf\nn Zahn O\dualand\nn Zaldarriaga M;2006;ApJ;653;922

\bibitem{Metcalf06}
\rf\nnn Metcalf R B\dualand\nnnn White S D M;2007;MNRAS;381;447

\bibitem{Hilbert07}
\rf\nn Hilbert S, \nnn Metcalf R B\multiand\nnnn White S D M;2007;MNRAS;382;1494

\bibitem{gsm}
\rf\nn {de Oliveira-Costa}, \nn Tegmark M, \nnn Gaensler B M, \nn Jonas J, \nnn Landecker T L\multiand\nn Reich P;2008;MNRAS;388;247

\bibitem{21cm}
\rf\nn Wang X, \nn Tegmark M, \nn Santos M\multiand\nn Knox L;2006;ApJ;650;529 

\bibitem{Morales06}
\rf\nnn Morales M F, \nnn Bowman J D\multiand\nnn Hewitt J N;2006;ApJ;648;767

\bibitem{Jelic08}
\rf\nn Jelic V {\etal};2008;MNRAS;389;1319

\bibitem{BowmanThesis}
\rn\nnn Bowman J D, MIT Ph.D.~thesis (2007)

\bibitem{Bowman08}
\rfprep\nnn Bowman J D, \nnn Morales M F\multiand\nnn Hewitt J N;2008;{arXiv:0807.3956 [astro-ph]}

\bibitem{21cmps}
\rfprep\nn Liu A, \nn Tegmark M\multiand\nn Zaldarriaga M;2008;{arXiv:0807.3952 [astro-ph]}


\bibitem{KolbTurnerBook}
\rfbook\nnn Kolb E W\dualand\nnn Turner M S;1990;The Early Universe;Addison Wesley;{New York}

\bibitem{LSST}
\rfprep\nn Ivezic Z {\etal};2008;{arXiv:0805.2366 [astro-ph]}



\end{thebibliography}
\end{document}